\pdfoutput=1
\documentclass[conference]{IEEEtran}
\IEEEoverridecommandlockouts
\usepackage{cite}
\usepackage{amsmath,amssymb,amsfonts}
\usepackage{algorithmic}
\usepackage{graphicx}
\usepackage{textcomp}
\usepackage{xcolor}

\usepackage{fancyhdr}
\usepackage[hyphens]{url}
\usepackage{comment}
\usepackage{multirow}

\usepackage{comment}
\usepackage{algorithmic}
\usepackage[ruled,linesnumbered,noend]{algorithm2e}

\usepackage[bookmarks=true,breaklinks=true,letterpaper=true,colorlinks,linkcolor=blue,citecolor=blue,urlcolor=black]{hyperref}

\usepackage{listings}
\newcommand{\squishlist}{
 \begin{list}{$\bullet$}
  { \setlength{\itemsep}{0pt}
     \setlength{\parsep}{0pt}
     \setlength{\topsep}{3pt}
     \setlength{\partopsep}{0pt}
     \setlength{\leftmargin}{1.5em}
     \setlength{\labelwidth}{1em}
     \setlength{\labelsep}{0.5em} } }
     
\newcommand{\squishnums}{
 \begin{list}{$\bullets$}
  { \setlength{\itemsep}{0pt}
     \setlength{\parsep}{3pt}
     \setlength{\topsep}{3pt}
     \setlength{\partopsep}{0pt}
     \setlength{\leftmargin}{1.5em}
     \setlength{\labelwidth}{1em}
     \setlength{\labelsep}{0.5em} } }

\newcommand{\squishlisttwo}{
 \begin{list}{$\bullet$}
  { \setlength{\itemsep}{0pt}
     \setlength{\parsep}{0pt}
    \setlength{\topsep}{0pt}
    \setlength{\partopsep}{0pt}
    \setlength{\leftmargin}{2em}
    \setlength{\labelwidth}{1.5em}
    \setlength{\labelsep}{0.5em} } }

\newcommand{\squishend}{
  \end{list}  }

\newcommand{\TODO}[1]{\textcolor{magenta}{TODO: #1}}

\newcommand{\union}{Union\xspace}

\def\BibTeX{{\rm B\kern-.05em{\sc i\kern-.025em b}\kern-.08em
    T\kern-.1667em\lower.7ex\hbox{E}\kern-.125emX}}
\begin{document}

% circit / taco / mapping / tensor
\title{Union: A Unified HW-SW Co-Design Ecosystem in MLIR for Evaluating Tensor Operations \\on Spatial Accelerators\\
}

\begin{comment}
The paper must have an abstract under 300 words.
The paper must be original material that has not been previously published in another conference or journal, nor is currently under review by another conference or journal. You may submit material presented previously at a workshop without copyrighted proceedings.
The submission is limited to eleven (11) pages in the IEEE 8.5” x 11” (US letter size paper) Two-Column Format using 10pt font, with no more than 6 lines per inch. This page limit applies to all content NOT INCLUDING references, and there is no page limit for references. 
\end{comment}
\author{\IEEEauthorblockN{Geonhwa Jeong\IEEEauthorrefmark{1}, 
Gokcen Kestor\IEEEauthorrefmark{3} \textsuperscript{\textsection}, Prasanth Chatarasi\IEEEauthorrefmark{2} \textsuperscript{\textsection},  Angshuman Parashar\IEEEauthorrefmark{6}, \\
Po-An Tsai\IEEEauthorrefmark{6}, Sivasankaran Rajamanickam\IEEEauthorrefmark{5}, Roberto Gioiosa\IEEEauthorrefmark{3}
and Tushar Krishna\IEEEauthorrefmark{1} 
    % \IEEEauthorblockA{\IEEEauthorrefmark{1}\textit{Sandia National Laboratories}, \IEEEauthorrefmark{2}\textit{Georgia Institute of Technology}}
    \IEEEauthorblockA{\IEEEauthorrefmark{1}\textit{Georgia Institute of Technology}, geonhwa.jeong@gatech.edu, tushar@ece.gatech.edu}
    \IEEEauthorblockA{\IEEEauthorrefmark{3}\textit{Pacific Northwest National Laboratory}, 
    \{gokcen.kestor, roberto.gioiosa\}@pnnl.gov}
        \IEEEauthorblockA{\IEEEauthorrefmark{2}\textit{IBM Research}, 
    \ prasanth@ibm.com}
    \IEEEauthorblockA{\IEEEauthorrefmark{6}\textit{NVIDIA}, 
    \{aparashar, poant\}@nvidia.com}
    \IEEEauthorblockA{\IEEEauthorrefmark{5}\textit{Sandia National Laboratories}, 
    \ srajama@sandia.gov}
    %\IEEEauthorblockA{\IEEEauthorrefmark{3}\textit{Intel Labs}, Bangalore, India
    %\{sudarshan.srinivasan, dipankar.das\}@intel.com}
    }
    }

%%%%%%%%%%%%% comment out for double blind review %%%%%%%%%%%%%%%%%
% \and
% \and
% \and
% \and
% \and

\maketitle
\begingroup\renewcommand\thefootnote{\textsection}
\footnotetext{Joint second authors}
\endgroup

\pagestyle{plain}
\begin{abstract}
To meet the extreme compute demands 
for deep learning across commercial and scientific applications, 
dataflow accelerators 
are becoming increasingly popular.
While these ``domain-specific" accelerators are not fully programmable like CPUs and GPUs,
they retain varying levels of flexibility 
with respect to data orchestration, i.e., dataflow 
and tiling optimizations to 
enhance efficiency. 
There are several challenges when designing new algorithms and mapping approaches to execute the algorithms for a target problem on new hardware. 
Previous works have addressed these challenges individually. 
To address this challenge as a whole, in this work, we present a HW-SW co-design ecosystem for spatial accelerators called \union \footnote{\url{https://github.com/union-codesign/union}} within the popular MLIR compiler infrastructure. 
Our framework allows exploring different algorithms and their mappings on several accelerator cost models.
\union also includes a plug-and-play library of accelerator cost models and mappers which can easily be extended. 
The algorithms and accelerator cost models are connected via a novel mapping abstraction that captures the map space of spatial accelerators which can be systematically pruned based on constraints from the hardware, workload, and mapper. 
We demonstrate the value of \union for the community with  several case studies which examine offloading different tensor operations (CONV/GEMM/Tensor Contraction) on diverse accelerator architectures using different mapping schemes.

%targeting diverse accelerator architectures such as systolic arrays (e.g., Google TPU), multi-chiplet spatial arrays (e.g., NVIDIA Simba) and fully-flexible CGRAs (e.g., Georgia Tech's MAERI).
%offloading different kinds of operators (CONV/GEMM/Tensor Contraction) on diverse accelerator architectures (rigid arrays / fully-flexible / multi-chiplet)
% Since layer shapes and dimensions vary significantly across models, 
% it is critical to determine how to map 
% the model efficiently onto the target accelerator.
% Still, it is cumbersome to find the optimal mapping from high-level to the accelartor, and people have developed mapper and cost-model in the ad-hoc manner.
% In this paper, we propose UNION, a unified ecosystem for evaluating DNN operators on various spatial architectures.
% We decouple various layers for hardware-software co-optimization so that computer architects, compiler researchers, and domain experts can solely focus on what they want to optimize.
% Using UNION, we show that different mappings can specify very different data allocations resulting in different performance and energy consumption.
% UNION provides efficient systematic method for DNN accelerator research including domain experts, compiler researchers and computer architects.
\end{abstract}


\section{Introduction}

Numerous custom ASIC accelerators have emerged in the recent past to effectively exploit massive parallelism and locality in the Machine Learning (ML) applications.
The most popular examples, such as TPU~\cite{tpu_isca17}, xDNN~\cite{xDNN-web}, RAPID~\cite{DBLP:conf/vlsic/FleischerSZSOSC18}, are based on the systolic arrays. 
There are also more advanced forms including NVDLA~\cite{nvdla},  Eyeriss~\cite{eyeriss_isca16},  ShiDianNao~\cite{shidiannao_isca15} and MAERI~\cite{maeri_asplos18}.
These accelerators have demonstrated lower runtime and higher energy efficiency relative to existing popular architectures such as multi-core CPUs and many-core GPUs~\cite{tpu_isca17}.
The main architectural features that distinguish these ``spatial'' accelerators from CPUs and GPUs are parallelism using hundreds to thousands of processing elements (PEs), efficient communication using a fast network-on-chip (NoC) to connect those PEs, and aggressive data reuse using private/shared scratchpad buffers with efficient scheduling. 
The success of these accelerators within the context of ML draws researchers' attention to using these accelerators in other compute-intensive domains as well, such as High Performance Computing (HPC) applications.
On the other hand, the accelerators are evolving rapidly with novel designs to support new application targets or to provide better performance. 
Comparing all those novel designs and understanding whether they can be good solutions to a specific algorithm/workload have become incredibly difficult for computer architects and compiler researchers. %especially in the early stages when multiple design choices are still open.
{\it Hence, there is a strong need for a flexible, composable, and reusable framework for evaluating new algorithms, their mappings on new spatial accelerator architectures.}

There are three critical components, algorithm/workload, mapping, and hardware for such an ecosystem. 
In the previous works, these are tightly coupled to each other.  
For example, 
%current simulators/analytical cost models are specific to a hardware or use existing
simulators~\cite{scalesim_ispass20, stonne_20} and analytical cost models~\cite{maestro_micro19, timeloop_ispass19} %templated hardware to estimate the performance of fixed set of tensor operators on 
are focusing on a limited set of accelerators.
They are also tightly coupled to a set of tensor operations as their inputs.
%to minimize the design-to-test time.
% Recent works have proposed analytical cost models~\cite{maestro_micro19, timeloop_ispass19} and simulators~\cite{scalesim_ispass20, stonne}. for evaluating spatial accelerators. 
%However, these cost models often have restrictions in modeling input computations and hardware micro-architectural features.
% Different cost models target a different set of accelerators. 
%If a user wants to evaluate tensor computations of different domains, the user needs to switch to different frameworks back and forth, which introduces non-trivial engineering overhead due to the lack of a unified ecosystem for evaluating accelerators.
%Moreover, users often need to re-implement benchmarks and kernels in a way that can be used to drive the framework, e.g., OpenMP, custom pragmas, or languages, PyTorch.
New high level interfaces or new algorithms sometimes require intrusive changes to the cost models.
In this work, we develop \emph{unified abstractions} in order to design a modular framework and mitigate the aformentioned problems.
The workload inputs for cost models vary depending on the cost models as well.
State of the art cost models require users to translate the operation in a specific format for the cost model~\cite{timeloop_ispass19} or translate a coarse-grained operation into fine-grained operations that the cost model understands~\cite{maestro_micro19}. 
Since this process is different depending on the frameworks, it requires manual efforts by users, which can be error-prone and tedious. 
A \textit{unified workload abstraction} for the cost models that we are presenting would get rid of this inefficiency.
% by implementing such an abstraction within a framework such as MLIR. 
%In addition to diverse workload files, 
The current cost models also differ in the mapping abstractions. For example, MAESTRO~\cite{maestro_micro19} uses data-centric mapping, Interstellar~\cite{interstellar_asplos20} uses Halide scheduling, and Timeloop~\cite{timeloop_ispass19} uses loop-nest mapping.
These abstractions have different strengths and limitations in expressing all possible mappings of various tensor computations and estimating cost metrics for these mappings on a new accelerator.
%Furthermore, these cost models are often developed with mapping space exploration tools to find efficient mappings on a target spatial architecture through various techniques such as exhaustive search, random sampling, heuristic-based pruning, and genetic algorithm.
The existing mappers~\cite{marvel, dmazerunner, gamma_iccad20, interstellar_asplos20, timeloop_ispass19, Flash2021, loma2021}, which find optimal mappings for the target workload and accelerator, are tightly dependent on their cost models due to the different mapping representations.
This limits the interoperability and reusability of the mappers even though conceptually mappers could be used among different cost models if they use a \textit{unified mapping abstraction}.
Finally, a \textit{unified hardware abstraction} is needed to represent a broad set of accelerators with diverse interconnects and memory hierarchies~\cite{simba_micro19, maeri_asplos18,eyeriss_v2,buffets_asplos19} to explore future hardware designs.

This work introduces \union, a unified ecosystem to evaluate tensor operations on spatial accelerators while addressing the challenges mentioned above. 
The ecosystem is designed with unified abstractions at every level, starting from a tensor operation and its mapping description to hardware description.
These abstractions enable the usage of different mappers and cost models interchangeably. Also, these abstractions are generic enough to use for future cost models and mappers.
Our ecosystem leverages the recently introduced MLIR infrastructure~\cite{mlir_cgo21} to integrate with different high-level languages or frameworks, such as Tensorflow, PyTorch for ML, and COMET~\cite{mutlu2020comet} for HPC. 
To the best of our knowledge, \union is the first framework unifying multiple high-level frameworks for tensor computations, mappers, and cost models for spatial accelerators.
We believe that our work would reduce the burden of computer architects, compiler researchers, and algorithm designers with our unified abstractions and ecosystem. 
In summary, the contributions of this paper are listed below: 

\begin{itemize}
    \item We provide a plug-and-play unified ecosystem to quickly evaluate tensor operations in various domains such as ML and HPC on spatial accelerators leveraging the MLIR infrastructure.
    \item We introduce new unified abstractions to describe tensor operations and their mappings on spatial accelerators to integrate different mappers and cost models. This allows us to evaluate diverse tensor operations from HPC kernels and ML use cases.
    \item We introduce operation-level/loop-level analysis to identify operations to be evaluated with the target spatial accelerator using a cost model.
    \item We show how our framework can be used with various workloads using different mappers and cost models for the diverse set of accelerators, including flexible and chiplet-based ones. The studies provide an inspiration for the future co-design of tensor operations and spatial accelerators. 
    %\item Evaluation: regular conv2d, conv3d... DSL for TC.. different accelerator configurations.. sparse kernels...Dense + Sparse workloads
\end{itemize}

We believe \union framework could enhance the co-design opportunities between compiler researchers, algorithm developers, computer architects and simulation tool developers. 

%\prasanth{First level pass is finished on intro.}

\section{Background}
\label{sec:background}

\begin{table*}[!ht]
\centering
\caption{Comparison of our framework UNION with other existing frameworks.}
\begin{tabular}{|c|c|c|c|c|c|c|c|c|}
\hline
\textbf{Framework}    & \textbf{\begin{tabular}[c]{@{}c@{}}Target \\ hardware\end{tabular}}     & 
\textbf{\begin{tabular}[c]{@{}c@{}}Cost\\ models\end{tabular}}                                          & 
\textbf{Mappers}                                                                       & \textbf{\begin{tabular}[c]{@{}c@{}}Operation\\ abstraction\end{tabular}}                   & \textbf{\begin{tabular}[c]{@{}c@{}}Mapping \\ abstraction\end{tabular}} & \textbf{\begin{tabular}[c]{@{}c@{}}Hardware\\ abstraction\end{tabular}} & \textbf{\begin{tabular}[c]{@{}c@{}}Integration with\\ high-level \\ frameworks\end{tabular}}  &
\textbf{\begin{tabular}[c]{@{}c@{}}Target\\ usecase\end{tabular}} \\
\hline
\textbf{AutoSA~\cite{autosa_fpga21}}       & Systolic                                                                & Custom                                                        & Custom                                                                                 & 
%Polyhedral (SCoP)                                                                                &
\begin{tabular}[c]{@{}c@{}}Polyhedral  \\ models\end{tabular} &

\begin{tabular}[c]{@{}c@{}}Space-Time \\ projections\end{tabular}       & \begin{tabular}[c]{@{}c@{}}Custom \\ format\end{tabular}                & N/A & Co-design                                                                                          \\ \hline
\textbf{MAESTRO~\cite{maestro_micro19}}      & Spatial                                                                 & Generic                                                       & \begin{tabular}[c]{@{}c@{}}Marvel, \\ GAMMA\end{tabular}                               & 
%\begin{tabular}[c]{@{}c@{}}Operator parameters\\ specific to \\ CONV2D, GEMM\end{tabular}        &
\begin{tabular}[c]{@{}c@{}}Fixed  \\ operations\end{tabular} &
\begin{tabular}[c]{@{}c@{}}Cluster-target  \\ Data-centric\end{tabular} & \begin{tabular}[c]{@{}c@{}}3-level \\ accelerators\end{tabular}         & \begin{tabular}[c]{@{}c@{}}Custom parser \\ from TF, PyTorch \\ (ML)\end{tabular} & Co-design            \\ \hline
\textbf{Timeloop~\cite{timeloop_ispass19}}     & Spatial                                                                 & Generic                                                       & \begin{tabular}[c]{@{}c@{}}Mind Mapping,\\ Random-based,\\ Brute-force\end{tabular} & 
%\begin{tabular}[c]{@{}c@{}}Perfectly affine \\ nested loops with \\ no conditionals\end{tabular} &
\begin{tabular}[c]{@{}c@{}}Nested  \\ loops\end{tabular} &

\begin{tabular}[c]{@{}c@{}}Memory-target  \\ Loop-centric\end{tabular}  & Hierarchical                                                            & \begin{tabular}[c]{@{}c@{}}Custom parsers \\ from TF (ML)\end{tabular} & Co-design                       \\ \hline
%%%%%%%%%%%%%%%Interstellar
\textbf{Interstellar~\cite{interstellar_asplos20}} & Spatial                                                                 & Generic                                                       & Heuristics                                                                             & 
%\begin{tabular}[c]{@{}c@{}}Operator parameters \\ specific to \\ CONV2D, GEMM\end{tabular}       & 
\begin{tabular}[c]{@{}c@{}}Fixed  \\ operations\end{tabular} &
%Halide scheduling                                                       & 
\begin{tabular}[c]{@{}c@{}}Halide  \\ scheduling\end{tabular} &
\begin{tabular}[c]{@{}c@{}}3-level \\ accelerators\end{tabular}         & N/A         & Co-design                                                                                  \\ \hline
\textbf{XLA}          & TPU                                                                     & Custom                                                        & Custom                                                                                 & LHLO                                                                                             & LHLO                                                                    & \begin{tabular}[c]{@{}c@{}}Specific \\ to TPU\end{tabular}              & TF (ML)  & Compilation                                                                                    \\ \hline
%%%%%%%%%%%%%ZigZag
\textbf{ZigZag~\cite{zigzag2021}} 
& Spatial                                                                
& Generic                                                       
& Heuristics, LOMA
& 
%\begin{tabular}[c]{@{}c@{}}Operator parameters \\ specific to \\ CONV2D, GEMM\end{tabular}       & 
\begin{tabular}[c]{@{}c@{}}Nested  \\ loops\end{tabular} &
%Halide scheduling                                                       & 
\begin{tabular}[c]{@{}c@{}}Memory-target  \\ Loop-centric\end{tabular} &
\begin{tabular}[c]{@{}c@{}}Hierarchical\end{tabular}         
& N/A         
& Co-design                                                                                  \\ \hline
\textbf{XLA}          & TPU                                                                     & Custom                                                        & Custom                                                                                 & LHLO                                                                                             & LHLO                                                                    & \begin{tabular}[c]{@{}c@{}}Specific \\ to TPU\end{tabular}              & TF (ML)  & Compilation                                                                                    \\ \hline

%%%%%%%%%%%%%%TVM
\textbf{TVM~\cite{tvm_osdi18}}          & \begin{tabular}[c]{@{}c@{}}Specific \\ (e.g., VTA)\end{tabular} & %\begin{tabular}[c]{@{}c@{}}Generic \\ (Auto-TVM)\end{tabular} & 
Generic &
Annealing                                                                              & 
%TVM statement                                                                                    & 
\begin{tabular}[c]{@{}c@{}}TVM  \\ statements\end{tabular} &
\begin{tabular}[c]{@{}c@{}}TVM  \\ scheduling\end{tabular} &
\begin{tabular}[c]{@{}c@{}}Specific to \\ target\end{tabular}           & TF, ONNX (ML)  & Compilation                                                                                \\ \hline \hline
\textbf{Union}        & Spatial                                                                 & Generic                                                       & Unified                                                                                & 
%\begin{tabular}[c]{@{}c@{}}Perfectly affine \\ nested loops with \\ no conditionals\end{tabular} &
\begin{tabular}[c]{@{}c@{}}Nested  \\ loops\end{tabular} &

\begin{tabular}[c]{@{}c@{}}Cluster-target\\ Loop-centric\end{tabular}   & Hierarchical                                                            & \begin{tabular}[c]{@{}c@{}}TF (ML), \\ COMET (HPC)\end{tabular} & Co-design                              \\ \hline
\end{tabular}
\label{comparision}
\end{table*}
%\subsection{DNN Operations}

\subsection{Tensor Operations}
In this section, we discuss several key tensor operations across ML and HPC.
%that we evaluate using \union.

\textbf{Deep Neural Network (DNN) Models.}
Recently, DNN models are outperforming other ML conventional techniques in various domains.
Convolution layers and fully-connected 
layers form the bulk of most DNN models, with the former dominating computer vision models and the latter dominating Natural Language Processing (NLP) and recommendation models. 
From an acceleration perspective, the 2D convolution (CONV2D) and generalized matrix-multiplication (GEMM) operations are being widely used to represent these two layers respectively. 
The \autoref{alg:conv} describes the convolution operation using the loop nest representation.
Some accelerators such as TPU~\cite{tpu_isca17} use algorithmic transformations such as the \textit{im2col}~\cite{cudnn} to convert
CONV2D to the GEMM operation while others directly compute convolution operations.

% A number of DNN operations has been introduced, but hardwares 
% spend most of their time working on two notorious but crucial operations: convolution layer and fully connected layer. Usually, convolution layers are extensively used on computer vision tasks since it effectively captures the local feature using a number of filters.  Fully connected layers are widely used in recommendation models as a part of MLPs. Also, FC layers are also used at the end of the CNNs to map the feature map to the specific category.

%\subsection{HPC Operations}
\textbf{HPC Kernels.}
\begin{algorithm}[t]
% \scriptsize
%\footnotesize
\SetInd{0.45em}{0.45em}
\SetAlgoLined
\DontPrintSemicolon
\KwIn{ 
\textit{IA}: An input activation with $[n][c][x][y]$ \\
\textit{OA}: An output activation with $[k][c][x'][y']$ \\
\textit{F}: An array of filters with $[k][c][r][s]$ \\
\textit{stride}: Stride for sliding windows\\
\\
}
%dataflow\_candidates $\leftarrow$ {\scriptsize \textbf{get\_candidate\_LoopOrders\_ClusterSzs}(\textit{Arch}, \textit{P})};\\
%\textit{num\_LoopOrders} $\leftarrow$ {\scriptsize \textbf{get\_num\_LoopOrders}(dataflow\_candidates)};\\ \textit{num\_ClusterSzs} $\leftarrow$ {\scriptsize \textbf{get\_num\_ClusterSzs}(dataflow\_candidates)};\\

% \textit{num\_LoopOrders} $\leftarrow$ dataflow\_candidates.shape[0]; \textit{num\_ClusterSzs} $\leftarrow$ dataflow\_candidates.shape[1];\\

% \Comment*[r]{\textbf{\textcolor{blue}{num\_LoopOrders = candidate\_dataflows.shape[0]}}}
% \Comment*[r]{\textbf{\textcolor{blue}{num\_ClusterSzs = candidate\_dataflows.shape[1]}}}

\For{n $=$ 0 \KwTo N-1}{
    \For{k $=$ 0 \KwTo K-1}{
        \For{x $=$ 0 \KwTo (X-R) / stride}{
            \For{y $=$ 0 \KwTo (Y-S) / stride}{
                \For{c $=$ 0 \KwTo C-1}{
                    \For{r $=$ 0 \KwTo R-1}{
                        \For{s $=$ 0 \KwTo S-1}{
                            $xx = x \times stride + r $\\
                            $yy = y \times stride + s $\\
                            $OA[n][k][x][y] += IA[n][k][xx][yy] \times F[k][c][r][s]$
        
                        }
                    }
                }
            }
        }
    }
}

\caption{A loop nest for a CONV2D Operation}
\label{alg:conv}
\end{algorithm}
\begin{algorithm}[t]
% \scriptsize
%\footnotesize
\SetInd{0.45em}{0.45em}
\SetAlgoLined
\DontPrintSemicolon
\KwIn{ $A$: A 4D input tensor with $[d][f][g][b]$ \\
$B$: A 4D input tensor with $[g][e][a][c]$ \\
$C$: A 6D output tensor with $[a][b][c][d][e][f]$ \\
\\
}
%dataflow\_candidates $\leftarrow$ {\scriptsize \textbf{get\_candidate\_LoopOrders\_ClusterSzs}(\textit{Arch}, \textit{P})};\\
%\textit{num\_LoopOrders} $\leftarrow$ {\scriptsize \textbf{get\_num\_LoopOrders}(dataflow\_candidates)};\\ \textit{num\_ClusterSzs} $\leftarrow$ {\scriptsize \textbf{get\_num\_ClusterSzs}(dataflow\_candidates)};\\

% \textit{num\_LoopOrders} $\leftarrow$ dataflow\_candidates.shape[0]; \textit{num\_ClusterSzs} $\leftarrow$ dataflow\_candidates.shape[1];\\

% \Comment*[r]{\textbf{\textcolor{blue}{num\_LoopOrders = candidate\_dataflows.shape[0]}}}
% \Comment*[r]{\textbf{\textcolor{blue}{num\_ClusterSzs = candidate\_dataflows.shape[1]}}}

\For{a $=$ 0 \KwTo A-1}{
    \For{b $=$ 0 \KwTo B-1}{
        \For{c $=$ 0 \KwTo C-1}{
            \For{d $=$ 0 \KwTo D-1}{
                \For{e $=$ 0 \KwTo E-1}{
                    \For{f $=$ 0 \KwTo F-1}{
                        \For{g $=$ 0 \KwTo G-1}{
                            $C[a][b][c][d][e][f] += A[d][f][g][b] \times B[g][e][a][c]$
                        }
                    }
                }
            }
        }
    }
}

\caption{A loop nest for a TC Operation}
\label{alg:tc}
\end{algorithm}
%\geonhwa{@Gokcen, please add text here}
Tensor Contraction (TC) operations are generalization of matrix multiplications with arbitrary dimensions. They are popular in HPC domains including many scientific and engineering problems, such as  quantum chemistry and finite-element methods. 
For example, the perturbative triples correction in couple cluster CCSD(T)~\cite{ccsd_t} methods used in the NWChem computational chemistry framework~\cite{nwchem,nwchemex_chemrev2021} produces a 6D output tensor from two 4D inputs tensor.
The corresponding loop nest is shown in ~\autoref{alg:tc}.
%\tushar{@gokcen/roberto/siva: can you please add the loop nest for TC like we have for CONV2D?}
Tensor contractions are computationally intensive and dominate the execution time of many computational applications, thus many optimizations have been developed to improve the performance of executing these kernels. 
Traditional compilers mostly focus on optimizations such as loop fusions, loop tiling, and loop reordering. 
High-level Domain-Specific Language (DSL) compiler, instead, can take advantages from re-formulating tensor contractions in a form that is amenable for execution of heterogeneous devices. 
For example, the COMET compiler~\cite{mutlu2020comet}, a DSL compiler for dense and sparse tensor algebra for chemistry and graph analytics,  reformulates  tensor contractions by rewriting them with  equivalent transpose-transpose-GEMM-transpose (TTGT) expressions. 
The TTGT computation first flattens the tensors into matrices via explicit tensor transposition and reshape operations, then executes GEMM, and finally folds back the resulting matrix into the original tensor layout.
The main advantage of this re-formulation comes from leveraging highly efficient GEMM accelerators such as the NVIDIA tensor core~\cite{tensorcore} or other novel dataflow accelerators, such as the ones targeted in this work. 
These advantages usually overcome the additional transpositions and generally yield higher performance.
However, rebuilding the semantics of a tensor contraction from optimized loops is complicated. To achieve high performance on novel dataflow architectures, it is paramount that a compiler retains the semantics of the language operations throughout all the optimization steps, which explain why most of the novel dataflow accelerators proposed leverage high-level languages.

%General Matrix Matrix Multiplication (GEMM) is a old but still crucial kernel in HPC domain.

%Tensor Contraction is another popular HPC kernel, which is the generalization of GEMM with high dimensional tensors.

%Tensor Contraction can be solved directly by its loop nest implementation. 
%There is another popular method to solve tensor contraction using Transpose Transpose GEMM Transpose (TTGT) to leverage existing high-performance GEMM kernels.

%The outputs of tensor contractions can be derived in various ways. For example, one can convert it to a GEMM problem while another can directly solve the problem. 

\subsection{Multi-Level Intermediate Representation (MLIR)}

To bridge the semantic gap between high-level language and low level Intermediate Representations (IRs), we leverage the MLIR framework. MLIR has been proposed for both reusability and extensibility~\cite{mlir_cgo21} and allows intergration of multiple IRs with different level of semantics at the same time. 

%MLIR is designed to bring together different compiler tool-chains and IRs under a unified compiler infrastructures. 
Currently, many languages and libraries exist, including TensorFlow, Rust, Swift, and Julia, that rely on their own specific IR. On the other hand, multiple target architectures are emerging, especially in the Artificial Intelligence (AI) domain. Maintaining all these compiler frameworks and porting each of them to any new architecture are challenging tasks, which may limit the scope of each language to a limited number of target architectures.
The MLIR framework addresses this fragmentation problem by proposing a modular and reusable IR stack that sits in between the language representation and the architectural representation~\cite{mlir_cgo21}. In this way, architectural specific operations and types can be encapsulated in specific IRs, while sharing common operations, types, and optimizations across languages and target architectures.

MLIR also supports the compilation of high-level abstractions and domain-specific constructs while providing a disciplined and extensible compiler pipeline with gradual and partial lowering.
The design of MLIR is based on minimal fundamental concepts and most of the IRs in MLIR could be fully customized. 
Users can build domain-specific compilers and customized IRs, as well as combining with existing IRs, opting in to optimizations and analysis. 
The core MLIR concepts include the followings.
\begin{itemize}
    \item \textbf{Operations} are the units of semantics and model concepts from ``instructions" to ``functions" and ``modules". An operation always has an unique opcode. It takes arbitrary number of static single assignment (SSA) operands and produces results. It may also have attributes, regions, blocks arguments, and location information as well. 
    \item \textbf{Attributes} provide compile-time static information, such as integer constant values, string data, or a list of constant floating point values.
    \item \textbf{Values} are the results of an operation or block arguments, and a value always has a type defined by the type system. A \textit{type} contains compile-time semantics for the value. 
    \item \textbf{Dialects} consist of a set of operations, attributes and types which are logically grouped and work together. 
    \item \textbf{Regions} are attached to an instance of an operation to provide the semantics (e.g., the method of reduction in reduction operation).
\end{itemize}
Moreover, a region comprises a list of blocks, and a block comprises a list of operations.  Beyond the built-in dialect in MLIR system, MLIR users can easily define new dialects, types, operations, analysis or transformation passes and so on. This feature makes MLIR easily extensible. 

In this work, we leverage MLIR to decouple high-level language semantics, general optimizations and transformations, and architecture-specific mappings focusing on operations, attributes, and dialects.

\subsection{Spatial Accelerators}
\begin{figure}[t]
\centerline{\includegraphics[scale=0.4]{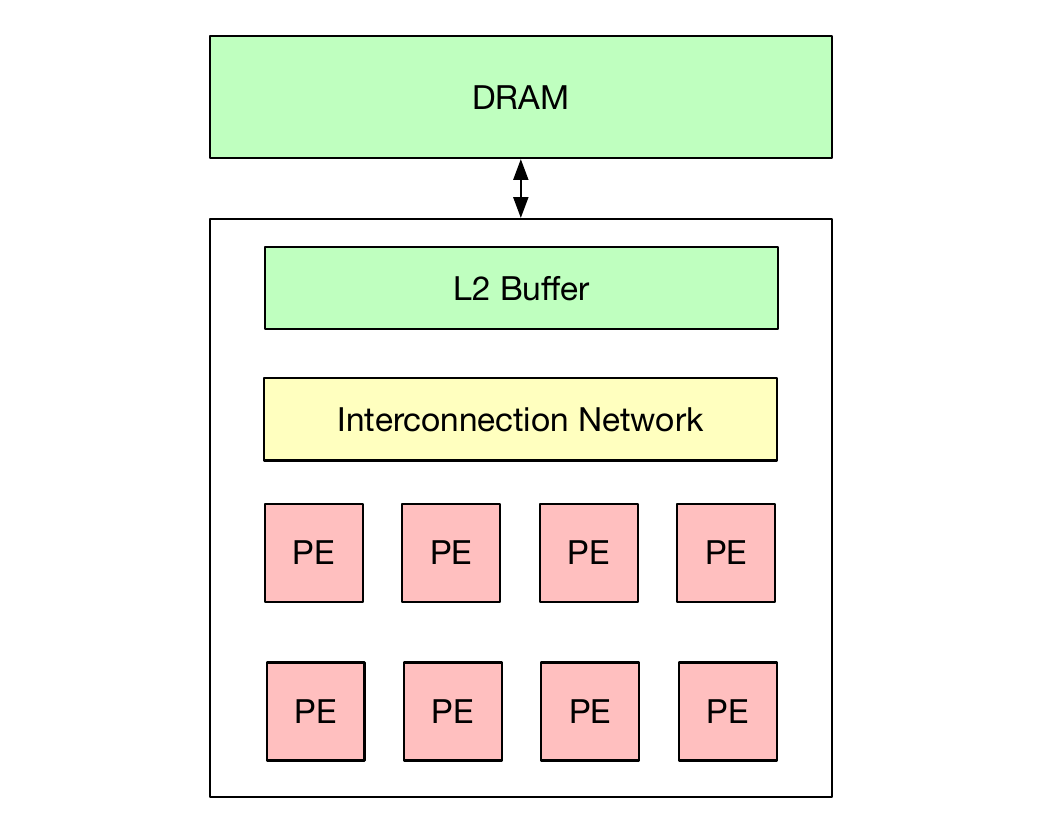}}
\caption{A simple spatial accelerator architecture with 8 PEs.
}
\label{fig:spatial_acc}
\end{figure}

To increase the compute throughput while achieving high energy-efficiency for DNN operations, various spatial accelerators have been proposed recently from both industry and academia. A simple spatial accelerator architecture composed of eight PEs with shared L2 buffer are shown in ~\autoref{fig:spatial_acc}.

\subsubsection{\textbf{Architecture}}
The spatial accelerators can be categorized into three groups based on their structure: rigid accelerators (e.g., TPU~\cite{tpu_isca17}, NVDLA~\cite{nvdla}, Eyeriss), flexible accelerators (e.g., Eyeriss\_v2~\cite{eyeriss_v2}, MAERI~\cite{maeri_asplos18}, SIGMA~\cite{sigma_hpca20}) and multi-chiplet accelerator (e.g., Simba~\cite{simba_micro19}). 
Unlike traditional architectures including CPUs and GPUs, spatial architectures use scratchpads as their intermediate buffers. 
Scratchpads are programmable so that the user can stage intermediate data tiles to maximize data reuse by properly \textit{mapping} the data at the right time at the right location. 

\subsubsection{\textbf{Cost Models}}
To quickly evaluate the performance and energy-efficiency of accelerators, the architecture community has been developing various cost models. 
Unlike CPUs and GPUs, where runtime contention for shared resources in the datapath and memory hierarchy can lead to non-determinism, accelerators can actually be modeled to the fairly accurate degree as their datapaths and memory hierarchies are tailored to 
the operation they are designed to accelerate. 
This allows accelerators to be modeled analytically without requiring cycle-level simulations.
Different cost models exist today 
in the community for 
modeling different kinds of accelerators at varying degrees of fidelity.
For e.g., SCALE-sim~\cite{scalesim_ispass20} models systolic arrays (e.g., Google TPU), 
MAESTRO~\cite{maestro_micro19} models spatial arrays with configurable aspect ratios~\cite{maeri_asplos18, eyeriss_v2}, Timeloop~\cite{timeloop_ispass19} can model hierarchical spatial arrays with complex memory hierarchies (e.g., partitioned buffers and buffer bypassing~\cite{buffets_asplos19}),  and Tetris~\cite{tetris_asplos17} can model 3D arrays.

% There can be a huge number of valid mappings for a single workload on the target spatial accelerator. To compare the performance of mappings without running each program using the mapping on the accelerator, there should be a way to statically evaluate the performance and/or energy of a single mapping on the target architecture.
% %theoretically. 
% %Similar to the methods that computer architects have used so far for CPUs and GPUs in the form of cycle-accurate CPU and GPU simulators, cycle-level simulators exist for accelerators~\cite{scalesim_ispass20, STONNE}.
% %for many of CPU and GPU simulators, there are a number of works which actually do the simulation cycle by cycle for accelerators~\cite{scalesim_ispass20, STONNE}. 
% Due to the long runtime for many simulations, another approach for evaluating a mapping on the target architecture is using an analytical cost models such as MAESTRO~\cite{maestro_micro19} and Timeloop~\cite{timeloop_ispass19}. 

\subsubsection{\textbf{Mappers}}
Using an accelerator cost model, one can estimate the performance of the program with a specific mapping on the target hardware. However, it is not straightforward to find the optimal mapping for a given workload and an architecture for two reasons.
First, the space of mappings can be extremely large~\cite{timeloop_ispass19} which makes exhaustive searches infeasible.
This has led to several mappers being developed to reduce the search time by pruning the search space or searching with efficient methods.
Marvel~\cite{marvel} proposes a decoupled approach
to decouple the off-chip map-space from the on-chip one, Timeloop~\cite{timeloop_ispass19} leverages sampling-based search methods, Interstellar~\cite{interstellar_asplos20} uses heuristic-based search, Mind Mapping~\cite{mindmapping_asplos21} develops 
a surrogate model to perform gradient-based search, and GAMMA~\cite{gamma_iccad20} uses genetic-algorithm based method to efficiently progress by leveraging the previous results.
This is currently an active area of research and we expect many more to come.
Next, defining the map-space can often be complex by itself since different operations and diverse hardware accelerators may impose constraints on the mappings that are feasible. 
This is the reason why the mappers described above are highly tied to specific cost models today, limiting interoperability. We discuss this further in the following section.

\subsection{Challenges with Existing Frameworks}

%Different cost models can model different accelerators. MAESTRO can model accelerators with different array shapes easily by changing the cluster size in the mapping file using the same hardware description file. Timeloop can model accelerators with partitioned buffers and buffer bypassing. 

The main challenge of the existing frameworks is that they have been developed in a tightly-coupled manner. For example, MAESTRO is a cost model which  estimates the performance of the hardware only when a mapping is given. 
Therefore, it does not find an optimal mapping for the hardware for a workload. GAMMA and Marvel are mappers which search for the optimal mapping for the target hardware/workload using MAESTRO as the cost model.
Since both GAMMA and Marvel are tied to MAESTRO, it is not possible to reuse mappers in GAMMA and Marvel using another cost model like Timeloop without having non-trivial engineering effort.
On the other hand, Timeloop includes both a cost model and a mapper. Similar to the previous example, it is not possible to use MAESTRO as the backend cost model using the Timeloop's mapper without significant engineering effort.
We summarize the comparison of our \union framework with prior frameworks in~\autoref{comparision}.
Since the goal of our work is to bring such Accelerator Design-Space Exploration tools under a unified framework, to the best of our knowledge, there is no such framework to compare directly with our approach.

Unfortunately, the lack of interoperability stifles innovation, since none of the mappers and cost models is perfect. Most new accelerator proposals develop 
new cost models for their design, but they are only able to 
demonstrate their efficiency for a few hand-optimized mappings. Similarly, researchers working on 
mapping/compilation for accelerators 
typically evaluate 
its efficiency on 
a specific accelerator for which they have access to the specific cost model (or real hardware).

This problem gets exacerbated as we move up the software stack 
since DNN model developers using high-level frameworks rely on very simple metrics like total number of Multiply-Accumulate (MAC) operations or the number of trainable parameters in their model to estimate the efficiency of the model which has been shown to be 
ineffective and often-times misleading~\cite{sze2020evaluate} as it 
loses all nuances related to the dataflow of the accelerator and data reuse capabilities.

We believe it is crucial to enable domain-experts, compiler experts, and computer architects 
to have an access to an end-to-end infrastructure 
that provides a 
library of plug-and-play mappers and cost models 
so that users can explore different options interchangeably, and focus on their specific 
research target (e.g., a new DNN model or a new mapper or a cost model for a new accelerator) without 
having to engineer or approximate the other parts.
Considering the features we discussed previously, MLIR can play a role as the right bridge for this effort.
%Since none of the mappers and cost models are perfect, one may need to change from a cost model to another or from a mapper to another. Therefore, it is critical to minimize the overheads so that a user can explore different options interchangeably. 

%we expect not to limit their power, but to expand their power by exploiting other works. Furthermore, researchers can focus on their target (i.e. a mapper design or a cost model design for new accelerator) without spending time for other layers.
\section{Overview of UNION}
\label{sec:union_overview}

\begin{figure}[t]
\centerline{\includegraphics[scale=0.4]{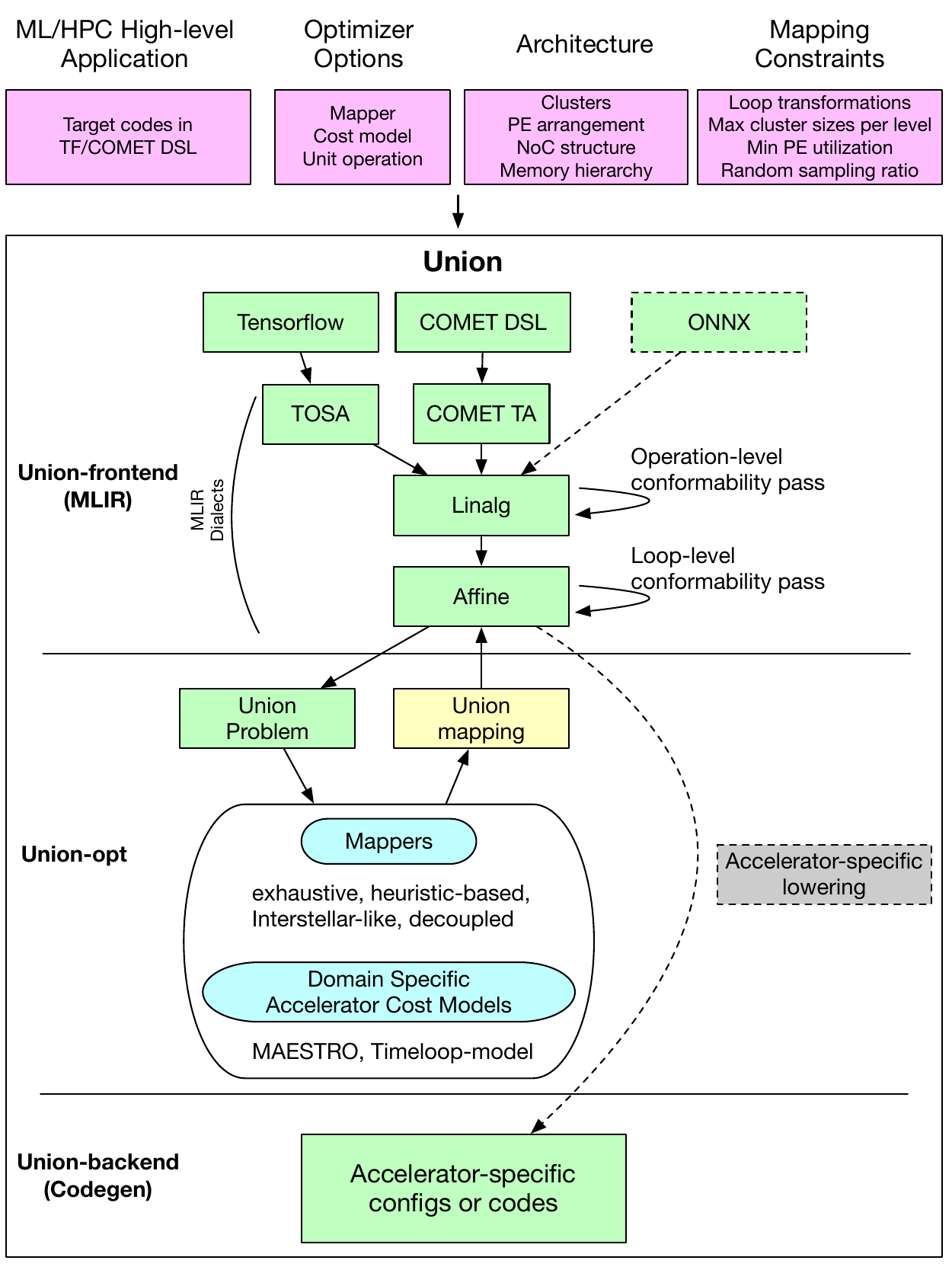}}
\caption{Union overview. The pink boxes indicate the inputs of \union while green boxes are showing how the codes are getting lowered. Rectangles and arrows with dotted lines are out of the scope of this paper. 
%\tushar{@Geonhwa: I have some suggestions on this figure. I think the ``UNION" box should start below the DSLs. And from TOSA to Affine we should have a curly bracket on the side and call it MLIR Dialects. I also feel like its not clear which of the input files get used at what stages.}
}
\label{fig:union_overview}
\end{figure}

In this section, we describe our framework, \union.
The overview of the framework is shown in \autoref{fig:union_overview}. 
A user will use \union by specifying workload in high-level language like TensorFlow or DSL, target hardware (with an architecture file and a mapping constraint file), and optimizer options including mapper type, cost model type and unit operation.
\union analyzes and lowers the given problem to a \union problem which is used for finding an efficient \union mapping that captures how the data should be tiled and delivered within the memory hierarchy.
The affine dialect annotated with a \union mapping can further be lowered to accelerator specific configurations to run the specific accelerators, but this is not the scope of this project and we leave it to the users who want to run their own accelerators.
% Commented by Tushar
%As mentioned before, the main problem with a diverse set of mappers and cost models is that they do not share a unified description for hardware, mappings, and problems.  Since they target to solve similar problem, many DNN mappers and cost models try to capture similar behavior and architecture by their own format, but it is not trivial to convert one from another. Furthermore, this become even worse if one needs to introduce a converter for each combination of different problem, mapper and cost model whenever a new problem or a mapper or a cost model gets introduced. 
One of the key contributions of 
\union is a set of abstractions for problem/hardware/mapping to unify different modules which will be presented in ~\autoref{sec:union_interface}.
Here, we introduce the overview of \union. 
%The details of the \union abstractions are explained in ~\autoref{sec:union_interface}. 
%In this work, we integrate two cost models, Timeloop~\cite{timeloop_ispass19} and MAESTRO~\cite{maestro_micro19}, for proof of concept, but \union can easily be integrated with other cost models.
%In the following sections, we show how our abstractions are used in \union-frontend and \union-opt. % in high level. 
%In this work, we implement and MAESTRO and Timeloop as the cost models since they use very different mapping representations and show how the proposed abstraction can be translated into their respective formats. 

% inputs of UNION

\subsection{Frontend: Using MLIR as a Bridge} \label{sec:frontend}

To demonstrate the composability and flexibility of our framework, we consider two high level DSLs which target very different application domains, TensorFlow for ML and COMET DSL for computational chemistry.

\subsubsection{\textbf{TensorFlow}} 
TensorFlow is one of the most famous open source platforms for machine learning.
Although several independent efforts exist to explore different ways of lowering TensorFlow code to mid-level MLIR dialects (such as linear algebra) including IREE~\cite{mlir-iree} and NpComp~\cite{mlir-npcomp}, we follow the Tensor Operator Set Architecture (TOSA) dialect approach~\cite{mlir-tosa} in this work. Moreover, current efforts, including ours, mostly focus the inference side and assume that the machine learning model has been already built and trained.
This approach is common on state-of-the-art DNN accelerators, such as the NVIDIA Deep Learning Accelerator (NVDLA)~\cite{nvdla}, where models are trained on GPUs and the NVDLA is used for inference.

We use a trained machine learning model using the standard execution flow on CPU, GPU, or TPU, which is saved as a graph. The graph is associated with a set of properties, including shape, types, and number of layers. Next, the generated graph is optimized and converted to its functional counterpart by removing some of the specific TensorFlow information, such as TensorFlow control regions and islands.
At this point, this graph can be converted to the TOSA dialect, which is a generic MLIR dialect for tensor algebra targeting machine learning applications.
The TOSA dialect is also the lowest domain-specific dialect in our framework.
As explained next, the rest of the compilation pipeline including mappers and cost models are shared across the various domains.

\subsubsection{\textbf{COMET DSL}}
The COMET compiler~\cite{mutlu2020comet,tian2021sparseComet} supports the COMET DSL for sparse and dense tensor algebra computations, focusing on computational chemistry kernels in NWChem and graph analytics.
The compiler is based on MLIR~\cite{mlir_cgo21} which performs a progressive lowering process to map high-level operations to low-level architectural resources.
It also includes a series of optimizations performed in the lowering process, and various IR dialects to represent key concepts, operations, and types at each level of the MLIR.
At each level of the IR stack, COMET performs different optimizations and code transformations. Domain-specific, hardware-agnostic optimizations that rely on high-level semantic information are applied at high-level IRs. These include reformulation of high-level operations in a form that is amenable for execution on heterogeneous devices (e.g., rewriting TC operations as TTGT) and automatic parallelization of high-level primitives (e.g., tiling for thread- and task-level parallelism). 
Currently, the compiler generates efficient code for traditional central processing unit (CPU) and GPU architectures as well as Verilog code for FPGAs.

\subsubsection{\textbf{Lowering to Linalg/Affine Dialect}}

Regardless of the language used for the original application, we lower the code down to the language-specific description of the application to frontend MLIR dialects, e.g., TensorFlow to TOSA or COMET DSL to COMET Tensor Algebra (TA) dialect.
This is shown in \autoref{fig:union_overview}.
%\tushar{@Geonhwa -- it seems like TA should be shown in Fig 2 then between COMET DSL and Linalg}
Next, we further lower from the domain-specific dialects to generic, language-independent constructs and operations, such as CONV2D and GEMM. 
In our framework, both TOSA and COMET TA dialects are lowered to a common Linear Algebra (Linalg) MLIR dialect.
At this stage, the IR is effectively decoupled from the original language and we can analyze the operations independently from the language. 
Depending on the accelerator cost model (as discussed next), the operations may be lowered further to Affine dialect for a loop-nest representation.

\textbf{Cost Model Dependent Conformability Passes.}
The next step needs to consider the requirements from the underlying cost models.
Here, \union transforms and annotates the generic IR obtained in the previous step with information that is necessary for the mapping design space exploration. 
The cost models we consider in this work targeting spatial accelerators have different constraints for the workloads that they can evaluate. For e.g., MAESTRO natively supports CONV2D and GEMM operations, where as Timeloop supports perfectly affine nested loops with no conditionals. 
Hence, our framework includes operation-level or loop-level conformability passes to check if the tensor operation is conformable to the underlying cost model for evaluation. 
These conformability passes embody different constraints (such as checking for specific operations or loop bounds~\cite{marvel}) of different cost models to determine whether it can be evaluated by the cost models.

%to integrate with mappers seamlessly.
%\tushar{point to a new figure/table that lists the conformability checks. We can highlight the ones needed for maestro vs timeloop. You can run it by Prasanth}

% \TODO{@Prasanth has lock to add text on conformability passes}

%\gokcen{are we completely removing passes loop analysis passes to determine the conformality of operator? Means that, we don't recognize the operator, it is automatically non-conformable?}

\subsection{An Optimizer for an Efficient Mapping: \union-opt}

\begin{figure}[t]
\centerline{\includegraphics[scale=0.15]{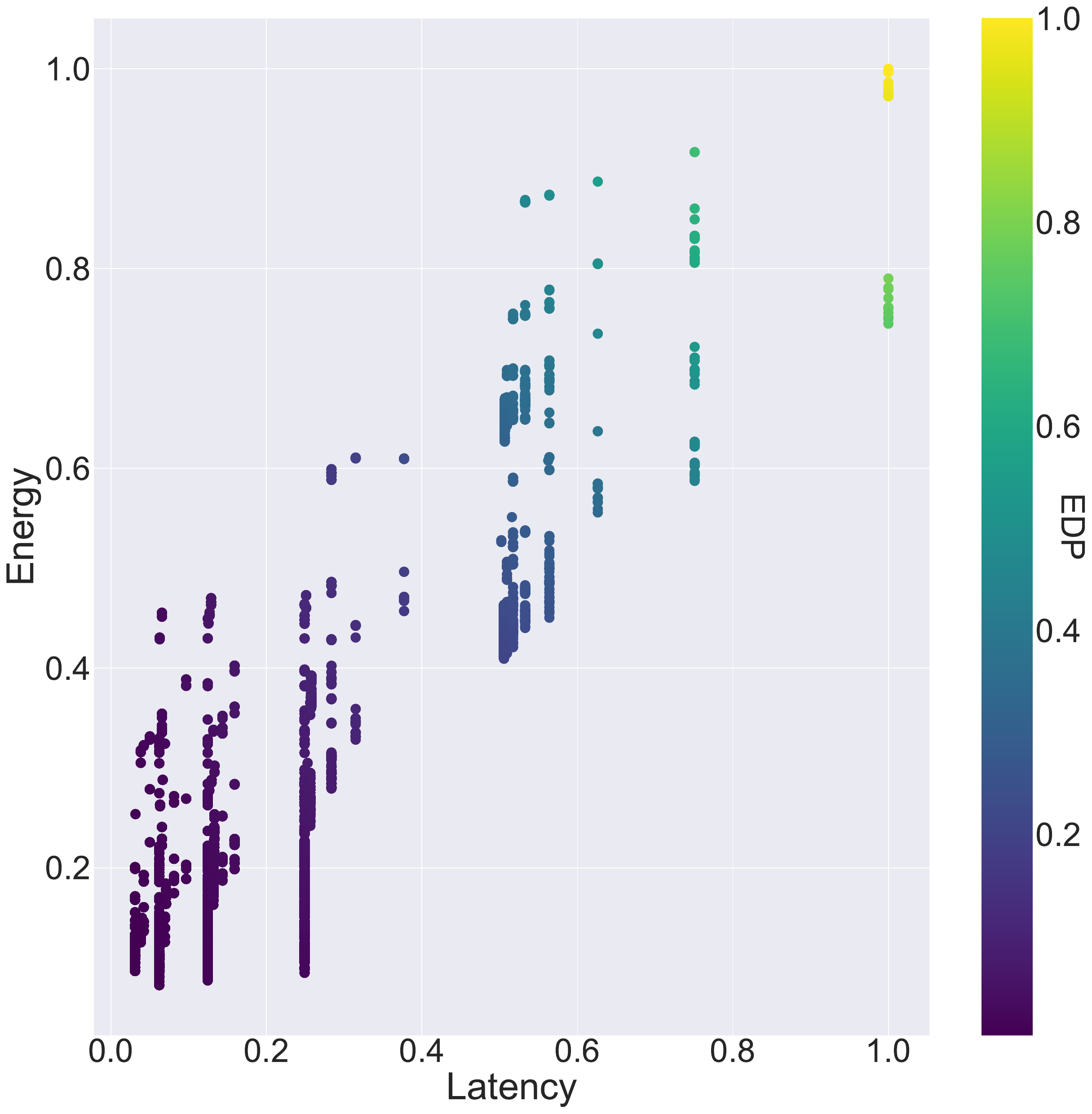}}
\caption{Normalized energy consumption and latency with EDP for different mappings of a layer from DLRM on a 3-level spatial architecture with 16$\times$16 PE array.}
\label{fig:mapping-space}
\end{figure}
After processing at the \union-frontend, the target problem is translated into an instance of \union problem abstraction (\autoref{sec:union_interface}).
The \union optimizer, \union-opt, searches for the efficient mapping of the problem based on the target metric such as latency, energy, Energy-Delay-Product (EDP).
To do so, \union-opt explores the \textit{map-space} for the given problem, architecture and constraint. 
%For example, how to run a problem in ~\autoref{fig:union_interface}(a) on the architecture in ~\autoref{fig:union_interface}(b) is determined by a mapping like ~\autoref{fig:union_interface}(d).
Different mappings can incur different PE utilization, data distribution, reduction, and data reuse affecting to the performance and energy efficiency.
~\autoref{fig:mapping-space} shows how the latency and energy consumption can vary for different mappings that \union explores for a layer from DLRM~\cite{dlrm} on a simple spatial architecture with 16$\times$16 PE array.
We will discuss more about how \union-opt can be used through the case studies in~\autoref{sec:case_studies}.
Since the mapping space for a simple problem can be extremely large due to the exponential and multiplicative characteristics of number of cases, it is inevitable to have efficient mappers other than exhaustive search~\cite{timeloop_ispass19}. 

\subsubsection{\textbf{Mappers}}
We currently integrate a few mappers in \union including exhaustive search, random sampling based search (from Timeloop~\cite{timeloop_ispass19}), 
decoupled approach (from Marvel~\cite{marvel}) and a few heuristic-based approaches. 
Users can add their own mappers and/or cost models by supporting our abstractions directly or adding converter from their format to our abstractions (\autoref{sec:union_interface}).

\subsubsection{\textbf{Domain-Specific Accelerator Cost Models}}
We currently implement Timeloop and MAESTRO as the cost models in \union to evaluate the mappings 
for proof of concept, but \union can easily be integrated with other cost models.

%\roberto{@geonhwa, maybe the next text should be moved to separate section?} 
MAESTRO~\cite{maestro_micro19} takes a high-level DNN operation such as CONV2D, GEMM, and DWCONV as an input problem. Therefore, whether the given problem is conformable or not depends on the high-level operation type. 
On the other hand, Timeloop~\cite{timeloop_ispass19} can take a fully nested loop which satisfies a few rules as an input problem. 
The fully nested loop should have affine indexes and every loop re-ordering should not change the result of the problem. 
Furthermore, each cost model assumes an unit operation for a PE such as two-operand MAC with certain data type.
To evaluate the performance of the given problem, the unit operation should be supported in its energy model.
For example, CONV2D can be used as an input problem for Timeloop since it can be described as a fully nested loop following the given rules as shown in \autoref{alg:conv} assuming that the energy model is configured with two-operand MAC as its unit operation.
Similarly, GEMM or Tensor Contraction can be evaluated using the Timeloop cost model since they all can be described as a fully nested loop following the given rules and using the two-operand MAC operation as its unit operation. Matricized tensor times Khatri-Rao product (MTTKRP) 
operation cannot be evaluated using Timeloop if its energy model is configured with two-operand MAC as its unit operation, but it can be done by changing three-operand-multiply-add as its unit operation and provide the necessary energy model. 
The backend of \union 
can be customized 
for generating configurations of individual accelerator targets. 
The backend is beyond the scope 
of this paper and part of our future work.

%Hardware DSE / Code generation

\begin{figure}[t]
\centerline{\includegraphics[scale=0.35]{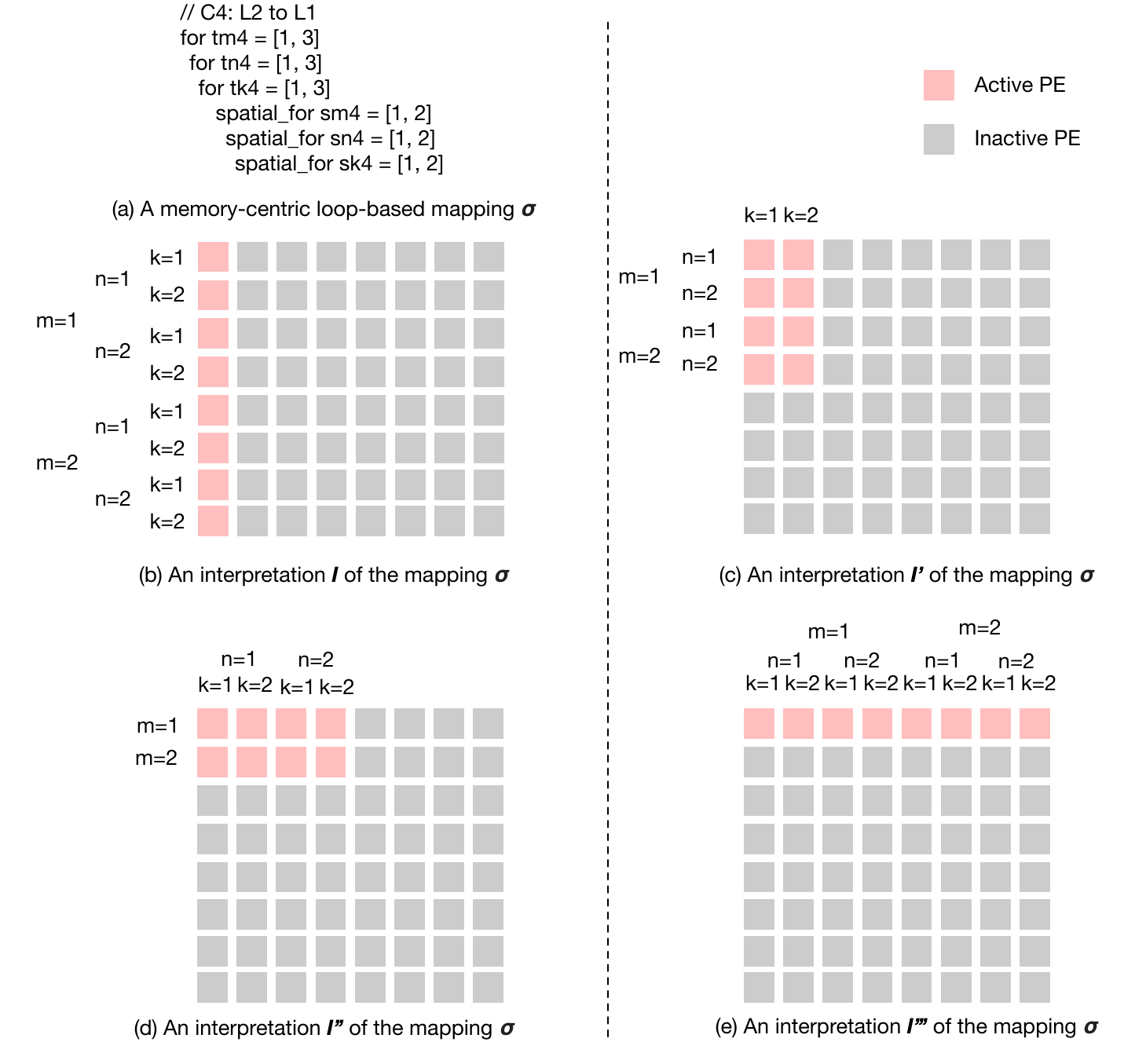}}
\caption{An example of a memory-target loop-centric mapping and its interpretations on a 8$\times$8 2D PE array.}
\label{fig:mem-centric-mapping}
\end{figure}

\begin{figure*}[t]
\centerline{\includegraphics[width=\textwidth]{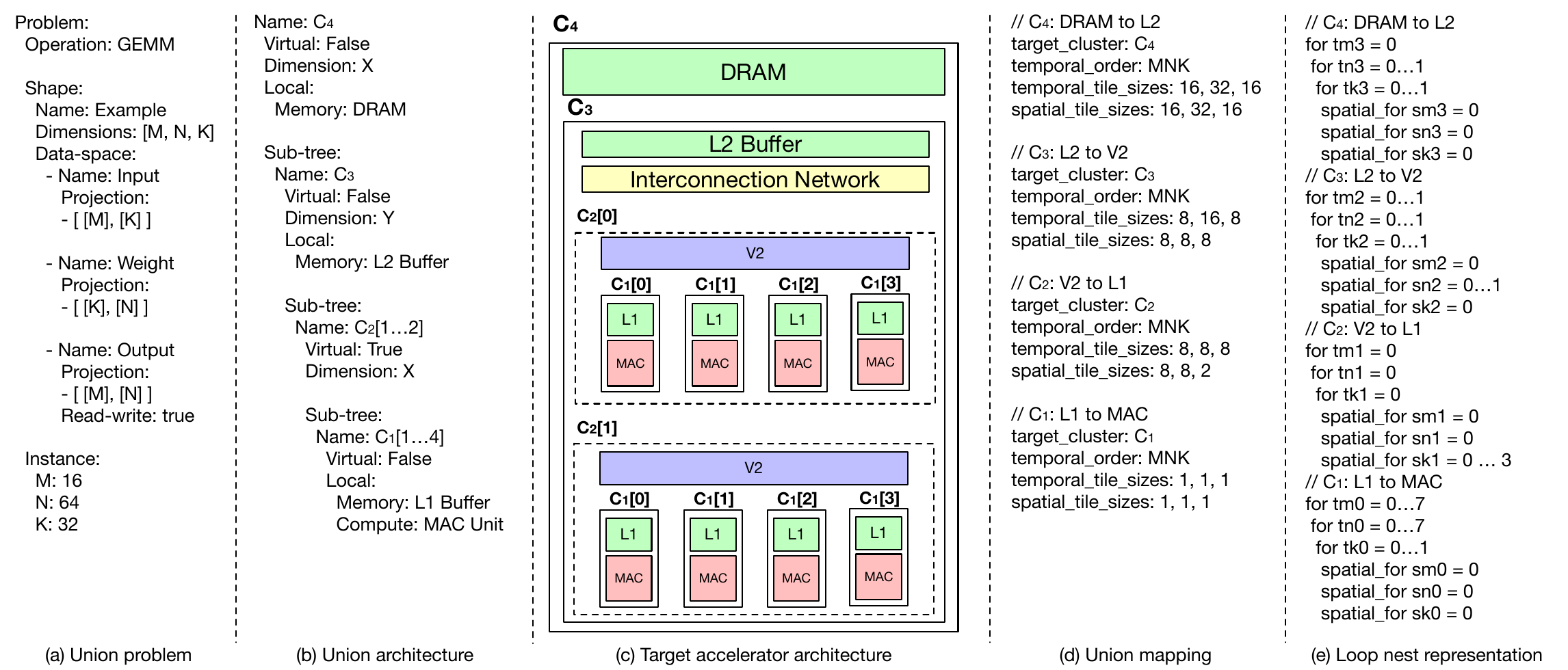}}
\caption{\union abstractions to describe a GEMM problem with a mapping on an accelerator which is composed of a simple 2D PE array. (a) describes a \union problem for a GEMM problem and (b) describes the \union architecture of the target architecture shown in (c). (d) shows a \union mapping that shows how to map the data to the architecture to run the GEMM problem. (e) represents the \union mapping in loop nest form.
%\prasanth{Potential Question: How do you capture PEs with tree-based interconnect using clusters?} 
}
\label{fig:union_interface}
\end{figure*}

\section{\union Abstractions for workload, architecture, and mapping}
\label{sec:union_interface}

Evaluating a mapped problem on a target spatial architecture requires abstractions between the architecture, mapping, and the workload.
Different frameworks that evaluates spatial accelerators have come up with different abstractions respectively to compare the performance and energy consumption of mappings of tensor operations on spatial architectures.
We first discuss some of their limitations and present the abstractions we developed for \union.

\subsection{Limitations of Current Mapping Abstractions}
%\TK{@Geonhwa -- it'll be good if this sub-section can be given some structure. First define what is memory-target loop-centric and Cluster-centric data-based. And then describe shortcomings in both } 

\subsubsection{\textbf{Memory-target Loop-centric Approach}}

Most of previous frameworks~\cite{timeloop_ispass19, interstellar_asplos20, marvel, dmazerunner, gamma_iccad20} use each hardware memory level as the target of a loop tiling level (i.e. tiling can only happen in each memory hierarchy level, such as between L2 and L1 buffers) to exploit temporal 
and spatial locality. 
%We call this \textit{memory-target loop-centric}.
%\subsection{Limitations of memory-target loop-centric approach}
~\autoref{fig:mem-centric-mapping} shows a memory-target loop-centric mapping $\boldsymbol{\sigma}$ and four different possible interpretations of such mapping on a 8$\times$8 2D PE array. 
Its loop nest representation is shown in ~\autoref{fig:mem-centric-mapping}(a) where \texttt{for} loops describe the temporal mapping while \texttt{spatial\_for} loops describe spatial mapping (i.e., parallel units).
Since the $\boldsymbol{\sigma}$ does not have the information about in which direction the problem dimensions is parallelized in physical spatial units, the mapping can be realized by all options in ~\autoref{fig:mem-centric-mapping}(b)-(e).
To circumvent the ambiguity the mapping representation, prior frameworks either assume certain implicit rules specific to the accelerators, or introduce extra annotations to indicate the mapping with spatial distribution and physical spatial axis.
Another limitation of such abstraction is that there is a 1-to-1 mapping between a tensor rank and physical spatial dimension in the memory-target representations. 
For example, memory-target abstraction cannot describe parallelizing the $M$ dimension onto both horizontal and vertical axis in the PE array.
Similarly, it is impossible to precisely describe a mapping which distributes $M$ and $N$ dimensions on horizontal axis and distributes $N$ and $K$ dimensions on the vertical axis using 
memory-target loop-centric mapping scheme.
Moreover, due to the hierarchical order between  \texttt{spatial\_for} loops, two iterators cannot change concurrently except at the loop bounds. Such limitation forbids a mapping which parallelizes different problem dimensions at the same time.
%The cluster-centric loop-based representation that we introduce in ~\autoref{sec:union_abstraction} effectively removes these restrictions and exposes more mappings to be described precisely.

\subsubsection{\textbf{Cluster-target Data-centric Approach}}
%On the other hand,
MAESTRO~\cite{maestro_micro19}  
introduces the notion of clusters. A cluster means a \textit{logical} group of PEs.
Instead of fixed hardware memory levels, MAESTRO targets each logical cluster level for tiling to explore more fine-grained tiling opportunities and remove the ambiguity caused by memory-target approach.
However, MAESTRO's mapping abstractions use a \emph{data-centric} notation which is not suitable to reason about using high-level compute-based abstractions, such as MLIR.
%However, it is not easy to explore 
%MAESTRO's mapping representations without understanding the directives and grammar of the MAESTRO DSL which introduces a new data-based notations which is very different from compute-based notations (i.e. loop-based approach) 
%which programmers are familiar with.
Moreover, MAESTRO assumes a \emph{fixed} accelerator architecture: a 2-level memory hierarchy with private L1 buffers and shared L2 buffer, so it is not possible to explore mappings for accelerator architectures with more complex memory hierarchy.
%Furthermore, none of the existing mappers based on MAESTRO~\cite{} does not explore the mapping space thoroughly

\union combines the best of both 
these approaches discussed above and 
introduces a \textit{logical cluster-target loop-centric approach}.
We adopt a cluster-target approach to be able to describe more mappings 
(addressing the shortcoming in Timeloop's representation) 
while still enabling a straightforward translation between our notation and loops from MLIR (addressing the shortcoming 
in MAESTRO's representation). Table~\ref{table:prior_abstration} shows the differences between prior abstractions and \union.

\begin{table}[t]
\centering
\caption{Comparison between prior abstraction and \union.}
\begin{tabular}{|r|c|c|} 
 \hline
 & \textbf{Hardware Memory-target} & \textbf{Logical Cluster-target} \\
 \hline
 \textbf{Data-centric} & N/A & MAESTRO~\cite{maestro_micro19} \\
 \hline
 \textbf{Loop-centric} & \begin{tabular}[c]{@{}c@{}}Timeloop~\cite{timeloop_ispass19}, \\ Interstellar~\cite{interstellar_asplos20} \end{tabular} & \union (\textbf{This work})\\
 \hline
\end{tabular} %\TODO{@geonhwa, can you fill in your citations?}}
\label{table:prior_abstration}
\end{table}

%In this work, we extend the abstractions using loop-based approach introduced in  Timeloop~\cite{timeloop_ispass19} to support broader tiling opportunities by adopting cluster-centric notions.

%\subsection{UNION abstraction} %\label{sec:union_abstraction}

\subsection{First Abstraction: From MLIR Dialects to a Problem Instance}
Common cost models support a set of workloads, defined in different levels (ex. operation level for MAESTRO~\cite{maestro_micro19} while loop level for Timeloop~\cite{timeloop_ispass19}). 
From the workload written in high-level language and the target cost model, \union-frontend extracts the information from both levels as an affine dialect with an operation annotation.
To handle the given problem with an affine dialect with an operation annotation, our abstractions includes loops, projections of the data spaces from array references, and operation type as shown in ~\autoref{fig:union_interface}(a)\footnote{inspired from Timeloop problem instance description}.
Loop iterators in the affine loop are set as dimensions and array references set each data in data-space with their projections. Finally, the size of each dimension is derived from the loop bounds.
The affine dialect is analyzed and re-organized to set dimensions, data-space, projection, and instance. 
We use the attribute Operation to indicate the operation (if given).
This abstraction captures both operation-level and loop-level information so that any cost model which supports one of them can be used. 
%\siva{Fig 5 needs better caption, also 5 (a) needs better explanation here. At least cite TimeLoop and say x, y, z are same as Timeloop.}

\subsection{Second Abstraction: Describing Architecture}

One of the key features of \union is to describe a logical spatial architecture instead of a fixed one.
We start with the hierarchical architecture abstraction used in the previous work~\cite{timeloop_ispass19} and extend it to describe the architecture in the logical cluster-target manner.
\autoref{fig:union_interface}(b) shows a \union architecture abstraction for the target spatial architecture illustrated in \autoref{fig:union_interface}(c).
The target architecture is composed of a 2D PE array, an L2 buffer shared across all PEs, and a private L1 buffer for each PE. 
We call the top cluster level in the $n$-level cluster architecture as $C_n$ in this paper. 
For example, in ~\autoref{fig:union_interface}(c), we call the outermost cluster level which has DRAM as its local memory as $C_4$ while the innermost cluster which has a L1 buffer as its local memory is called as $C_1$.

Various features can be specified in each cluster level such as compute, memory, and sub-clusters (and size of each sub-cluster).
We also add two new attributes in each cluster level, \textbf{Virtual} and \textbf{Dimension} in addition to the abstractions used in the previous work~\cite{timeloop_ispass19}.
The first attribute, Virtual, indicates whether the cluster has a dedicated physical memory or not.
The second attribute, Dimension, defines how the sub-clusters are laid in the physical dimension.
In the example architecture shown in \autoref{fig:union_interface}(c), the cluster at $C_4$ has DRAM as its memory and is composed of a single sub-cluster as $C_3$.
A cluster at $C_3$ has L2 buffer and is composed of two instances of $C_2$ which are laid in Y-axis. 
A cluster at $C_2$ is composed of four instances of $C_1$ which are laid in the X axis. 
Note that \textit{Virtual} is \textit{True} only for $C_2$ since $C_2$ does not have a dedicated memory. 
Instead, we draw $V_2$ in the figure which will always be bypassed since it is virtual (imaginary) buffer, but it provides a way to describe the intermediate tiling.
The innermost cluster, $C_1$, includes L1 buffer and a MAC unit. 
With \union architecture abstraction, one can describe how a multi-level clusters mapped on to multi-dimensional PE arrays. 
We assume that the parallelism can only be defined across sub-clusters.
For example, one can put another virtual cluster between $C_2$ and $C_1$ to exploit more fine-grained parallelism.
One can also describe partitioned buffer by introducing sibling clusters in the same cluster level, similar to the way how Timeloop~\cite{timeloop_ispass19} describes. 
Some architectures are limited to support certain loop orders depending on its dataflow such as input stationary, output stationary, weight stationary or row stationary~\cite{eyeriss_isca16}.
Those architectures can be realized by specifying the limitations in the constraint file which we discuss later.

\subsection{Third abstraction: Describing a Mapping between a Problem Instance and a Spatial Accelerator}
%The cluster-target loop-based representation that we introduce in ~\autoref{sec:union_abstraction} effectively removes these restrictions and exposes more mappings to be described precisely.

A mapping describes how a problem instance will be executed on a logical cluster-based architecture.
%A mapping defines which data will be mapped onto each storage level at different time. 
We propose a cluster-target mapping representation using loop-centric approach. 
Previous loop-based representations~\cite{timeloop_ispass19, interstellar_asplos20, marvel} describe the temporal/spatial behavior of tiles in each memory level while our proposed mapping describes the behaviors in each cluster level.
In our mapping, the parallelism across sub-clusters can be described at each cluster level.
Unlike the memory-target representations, one can describe tiling at a virtual cluster level even though this level does not have dedicated memory units.

\textbf{Semantics and characteristics.}
%We extend the hierarchical architecture abstraction introduced in Timeloop~\cite{timeloop_ispass19} using cluster-centric representation with \textit{Virtual} and \textit{Dimension} attributes. 
In our mapping abstraction, each tiling level explicitly targets a cluster, not memory, to cover broader mapping variants and remove ambiguity. 
An example \union mapping is shown in~\autoref{fig:union_interface}(d) and its loop nest representation is shown in~\autoref{fig:union_interface}(e).
\texttt{target\_cluster} defines the cluster level of the following tiling directives.
\texttt{temporal\_order} defines the temporal ordering between dimensions in the cluster level.
\texttt{temporal\_tile\_sizes} and \texttt{spatial\_tile\_sizes} defines the size of temporal and spatial tile for each dimension.

The spatial tile sizes defined in $(i+1)$th level cluster, $C_{(i+1)}$, can further be divided into sub-tiles in $C_i$. 
The tile can be divided into multiple time steps using temporal tiles in $C_i$. 
Each temporal tiles have the size as specified in \texttt{temporal\_tile\_sizes} in $C_i$
A temporal tile in $C_i$ can be divided into smaller spatial tiles and be spatially distributed into multiple instances of sub-clusters $C_{(i-1)}$. 
Therefore, the parallelism in $i$th level can be calculated by dividing temporal tile size by spatial tile size.
Note that we do not define \texttt{spatial\_order} in each cluster level. 
We change the semantic of \texttt{spatial\_for} so that it can change iterators concurrently in the same cluster level, inspired from MAESTRO data-centric notations~\cite{maestro_micro19}. 

Finally, \union introduces a few rules to check if a mapping is legal for the target logical architecture and the problem instance as shown in the following.
\begin{itemize}
    \item The mapping will be illegal if the spatial tile size of the problem dimension $d$ at $i$th cluster level is smaller than the temporal tile size of the same problem dimension $d$ at $(i-1)$th cluster level.
    \item The parallelism for the problem dimension $d$ at $i$th cluster level, which can be derived as $\frac{TT_d^i}{ST_d^i}$ should be equal to or smaller than the number of $(i-1)$th clusters in a $i$th cluster level.
    \item If a $i$th cluster is not virtual, the size of its memory should be as large as the memory sizes required by temporal tile sizes. $TT_d^i$ and $ST_d^i$ are the temporal and spatial tile size of problem dimension d in $i$th cluster respectively.
    \item The mapping should cover all the iteration vectors defined by the problem.
\end{itemize}

\begin{comment}
For $i$th cluster level, the number of possible temporal order choices, $C_{TO}^i$ is $(N_{dims})!$. 
For $i$th cluster level, the number of possible temporal tile size choices for a problem dimension $d$, $T_{(t, d)}^{i}$ is $T_{(t, d)}^{(i+1)}$. 
Therefore, the number of possible temporal tile sizes choices at $i$th cluster level,  
$C_{TT}^i$ is $\prod_{d \in D}{} T_{(T, d)}^{i}$.
Similarly, the number of possible spatial tile sizes choices at $i$th cluster level, 
$C_{ST}^i$ is $\sum_{d \in D}{} T_{(S, d)}^{i}$ assuming that only one problem dimension is distributed across the sub-clusters.
\end{comment}
\begin{figure}[t]
\centerline{\includegraphics[scale=0.24]{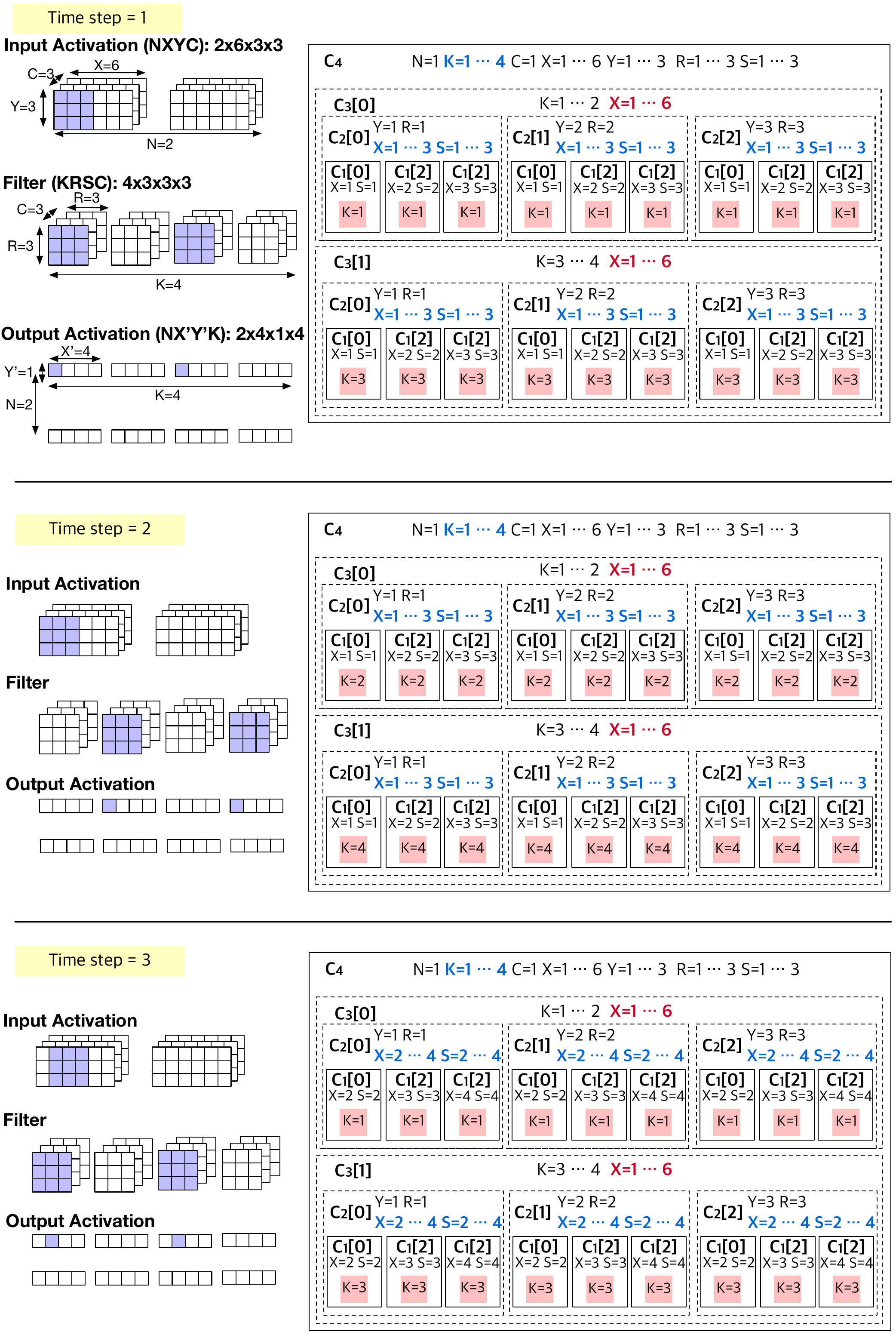}}
\caption{The visualization of a \textit{K\_YR\_XS} partitioned mapping for CONV2D using with 18 PEs for 3 time steps. A cluster containing DRAM, $C_5$, is not shown here. A red box indicates a MAC unit.}
\label{fig:maeri-visualization}
\end{figure}
\begin{figure}[t]
\centerline{\includegraphics[scale=0.37]{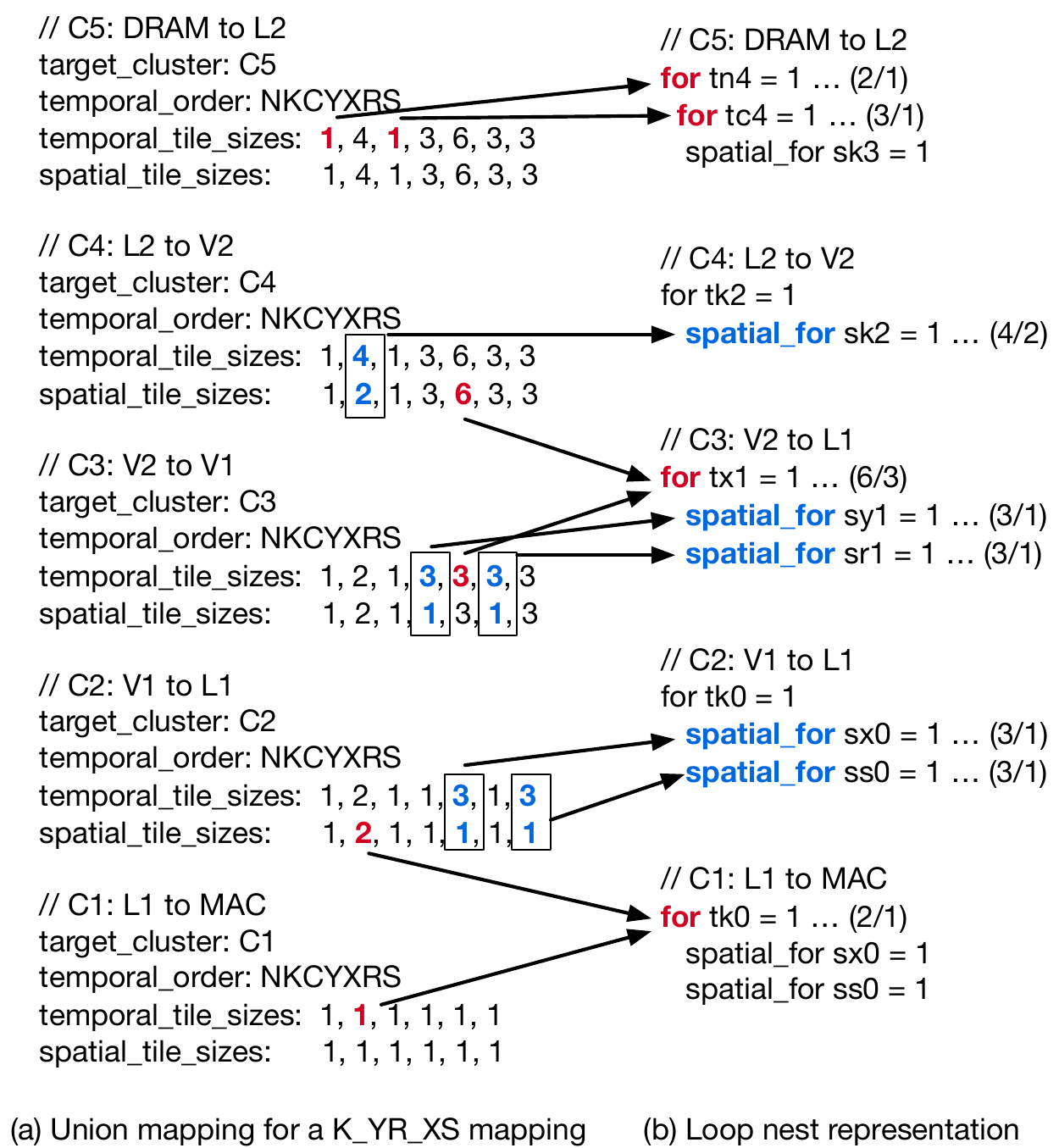}}
\caption{A \union mapping of a \textit{K\_YR\_XS} partitioned mapping and its loop nest representation. The order of dimensions in tile\_sizes is NKCYXRS.
%\prasanth{Are fig. 6 and fig. 7 connected? if so, I would put them in a single figure on two columns}
}
\label{fig:maeri-mapping}
\end{figure}

\textbf{Walk-through Example.} \autoref{fig:union_interface}(d) shows a \union mapping and \autoref{fig:union_interface}(e) describes the mapping using the loop nest representation.
A \union mapping can describe multi-level clusters of multi-dimensional PE arrays precisely and specify temporal and spatial tiling of data at each level.
%partial tiling (with extra temporal loops).
Also, one can distribute different problem dimensions at the same time in each cluster level over sub-clusters, i.e. there is no temporal ordering between  spatial\_fors in the same cluster level. 
%K_YR_XS Explanation
A complex example mapping for a small CONV2D operation using a flexible accelerator (such as MAERI~\cite{maeri_asplos18}) is illustrated in ~\autoref{fig:maeri-visualization} and the corresponding \union mapping and loop nest representation are shown in ~\autoref{fig:maeri-mapping}. 
In ~\autoref{fig:maeri-visualization}, the right column shows which data dimension values are mapped onto each cluster.
Clusters with solid lines ($C_4$ and $C_1$) have the dedicated memory in the level while clusters with dotted lines ($C_3$ and $C_2$) do not.
We use L2 and L1 to indicate the memory in $C_4$ and $C_1$ and V2 and V1 to indicate the virtual (imaginary) memory in $C_3$ and $C_2$.
The left column of ~\autoref{fig:maeri-visualization} visualizes the mapped elements in input activation, filter, and output activation. 
Purple elements are mapped onto MAC units at each time step.
%MAERI provides a flexible array of PEs that can be configured to spatially unroll and map tiles of several dimensions. 
In this mapping, dimension K is spatially distributed across $C_3$ clusters, Y and R are spatially distributed together across $C_2$, and X and S are spatially distributed together across $C_1$.
We call this mapping as a \textit{K\_YR\_XS} partitioned mapping to show the parallelism.
Each $C_2$ cluster is assigned for a row of a channel of a filter and a row of a channel of a input activation and each column in the row will be processed in the $C_1$ clusters concurrently.
Each $C_3$ cluster is assigned for a channel of a filter and the corresponding input activations.
As a result, inputs and outputs are reused between time step 1 and 2 while fetching different filters from the upper memory levels.
Between time step 2 and 3, a part of input activations are reused in MAC units and others are being fetched from the upper memory levels.

%\tushar{@Geonhwa - walk through this example in more detail, describing whats happening in each time step. use the K\_YR\_XS name we discussed}.

%a MAERI-like mapping shown in ~\autoref{fig:maeri-like} distributes X and S dimensions across different cluster0's.

%Using our cluster-centric loop-based representation, we can resolve the above limitations by exploiting more flexibility in the mapping.

%You can create multi-level clusters N-dimensional PE arrays and describe them precisely.
%Improve flexibility and expressibility
%It opens an opportunity to explore partial tiling.
%Expand mapping space (unsure whether this will improve performance)
%It opens an opportunity for inter-spatial-for and intra-spatial-for. (Change two dimensions simultaneously across different PEs)
%Expand mapping space 
\subsection{Constraint File}
In addition to \union abstractions, a user can also provide \emph{constraints} derived from a specific accelerator, such as feasible tile sizes, loop orders, parallelizing dimensions, and aspect ratio.
%\tushar{an example of a constraints file will be good to have. will complement fig 5.}
Such constraints provide the framework extra rules to eliminate illegal mappings and/or prune the mapping space for specific accelerators. For example, to describe a fully flexible accelerator like MAERI, the user will not provide constraint file to describe the hardware. On the other hand, a NVDLA-style~\cite{nvdla} architecture can be realized by having a constraint file that forces parallelization on dimensions C and K for a convolution operation with a fixed aspect ratio.
Furthermore, a user can set some constraints to prune the map space based on min/max PE utilization or specific loop orders or tile sizes that the user wants to explore.
\section{Case studies using \union}
\label{sec:case_studies}
\begin{table*}[t] 
\centering
\caption{Tensor contraction problems and the corresponding GEMM dimension sizes for TTGT}
\begin{tabular}{|c|c|c|c|}
\hline
\textbf{Name} & \textbf{Equation} & \textbf{Tensor Dimension Sizes} & \textbf{GEMM Dimension Sizes} \\
%\cline{2-4} 
\hline
\hline
intensli2 & C[a, b, c, d] = A[d, b, e, a] $\times$ B[e, c] 
& \begin{tabular}[c]{@{}c@{}} a = b = c = d = e = 64\\ a = b = c = d = e = 16\end{tabular} 
& \begin{tabular}[c]{@{}c@{}}M = 262144, N = 64, K = 64\\ M = 4096, N = 16, K = 16\end{tabular} \\
\hline
%intensli6 & C[a, b, c, d, e] = A[e, c, b, f, a] * B[f, d] & 64 & M=16777216 K=64 N=64\\
%& & 16 & M=65536 K=16 N=16\\
%ao2mo1 & C[a, b, c, d] = A[e, b] * B[a, e, c, d] & 64 &  M= 64 K= 64 N=262144\\
%& & 16 & M=16 K=16 N=4096\\
%ao2mo2 & C[a, b, c, d] = A[e, c] * B[a, b, e, d]; & 64 & M=64 K= 64 N=262144 \\
%& & 16 &  M=16 K= 16 N=4096\\
%ccsd3 & C[a, b, c] = A[a, c, d] * B[d, b] & 64 & M=4096 K=64 N=64\\
%& & 16 & M= 256 K=16 N=16\\
ccsd7 & C[a, b, c] = A[a, d, e, c] $\times$ B[e, b, d] 
& \begin{tabular}[c]{@{}c@{}}a = b = c = d = e = 64\\ a = b = c = d = e = 16\end{tabular} 
& \begin{tabular}[c]{@{}c@{}}M = 4096, N = 64, K = 4096 \\ M = 256, N = 16, K = 256 \end{tabular} \\
\hline
%ccsd9 & C[a, b, c, d] = A[a, e, b, f] * B[f, d, e, c] &  64 & M=4096 K=4096 N=4096\\
%& &  16 & M=256 K=256 N=256\\
ccsd-t4 & C[a, b, c, d, e, f] = A[d, f, g, b] $\times$ B[g, e, a, c] 
& \begin{tabular}[c]{@{}c@{}}a = b = c = d = e = f = g = 32\\ a = b = c = d = e = f = g = 16\end{tabular} 
& \begin{tabular}[c]{@{}c@{}}M = 32768, N = 32768, K = 32\\ M = 4096, N = 4096, K = 16\end{tabular} \\
\hline
%ccsd-t8 & C[a, b, c, d, e, f] = A[e, f, g, c] * B[g, d, a, b] & 32 & M=32768 K=32 N=32768\\
%&  & 16 & M=4096 K=16 N=4096\\
%ccsd-t12 & C[a, b, c, d, e, f] = A[g, e, a, b] * B[d, f, g, c] & 32 & M=32768 K=32 N=32768\\
%& & 16 & M= 4096 K=16 N=4096\\

%\multicolumn{4}{l}{$^{\mathrm{a}}$Sample of a Table footnote.}
\end{tabular}
\label{tab1}
\label{table:tc_workloads}
\end{table*}
\begin{table}[htbp] 
\caption{DNN Layer Dimensions Used in Evaluation}
\begin{center}
\begin{tabular}{|c|c|}
\hline
\textbf{Layer}&\textbf{Dimensions} \\
%\cline{2-4} 
\hline
ResNet50-1 & N=32 K=C=64 X=Y=56 R=S=1  \\
% conv_2_1_1
ResNet50-2 & N=32 K=C=64 X=Y=56 R=S=3  \\
% conv_2_1_2
ResNet50-3 & N=32 K=512 C=1024 X=Y=14 R=S=1  \\
% conv_5_1_1
DLRM-1 & N=512 NIN=1024 NON=1024  \\
% bottom_1
DLRM-2 & N=512 NIN=1024 NON=64  \\
% bottom_6
DLRM-3 & N=512 NIN=2048 NON=2048  \\
% top_10
BERT-1 & N=256 NIN=768 NON=768  \\
% bert 1-1
BERT-2 & N=256 NIN=3072 NON=768  \\
% bert 1-5
BERT-3 & N=256 NIN=768 NON=3072  \\
% bert 1-6

\hline
%\multicolumn{4}{l}{$^{\mathrm{a}}$Sample of a Table footnote.}
\end{tabular}
\label{tab1}
\end{center}
\vspace{-2mm}
\label{table:dnn_workloads}
\end{table}
\begin{table}[t]
\centering
\caption{Accelerator configurations}
\begin{tabular}{|c|c|c|c|c|c|}
\hline
\textbf{Type} & \textbf{\begin{tabular}[c]{@{}c@{}}\# of \\ PEs\end{tabular}} & \textbf{\begin{tabular}[c]{@{}c@{}}L1 Buffer\\ Size\end{tabular}} & \textbf{\begin{tabular}[c]{@{}c@{}}L2 Buffer\\ Size\end{tabular}} & \textbf{\begin{tabular}[c]{@{}c@{}}NoC\\ Bandwidth\end{tabular}} \\ \hline
Edge        & 256                                                           & 0.5 KB                                                            & 100 KB                                                            & 32 GB/s                                                        \\ \hline
Cloud       & 2048                                                          & 0.5 KB                                                            & 800 KB                                                            & 256 GB/s                                                                                                               \\ 
\hline
\end{tabular}
\label{table:hardware_config}
\end{table}

In this section, we show three case studies for algorithm exploration, mapping exploration, and hardware exploration to illustrate how \union can be used by domain experts, compiler researchers, and hardware architects, respectively.
We evaluate two types of accelerators, edge and cloud, as shown in~\autoref{table:hardware_config}. We assume 1GHz as the clock frequency and 8 bits for the wordsize with uint8 MAC units. 
In \union, we directly use Timeloop and MAESTRO, which are already validated against RTL for different existing accelerators. 
Thus, the validation of the performance numbers is dependent on the fidelity of the underlying cost models.

We choose tensor contractions from the TCCG benchmark suite~\cite{SpringerTCCG2018}, using the reference problem sizes.  The input sets are taken from different domains, including machine learning, molecular dynamics, and quantum chemistry. 
We use 16, 32, 64 as the Tensor Dimension Sizes (TDS) and assume that every dimension has the size as TDS in a TC problem instance.
We use a few representative DNN layers from the MLPerf benchmark including ResNet50 for computer vision, DLRM for recommendation, and BERT for natural language processing. 
We use N as a batch size, and K, C, X, Y, S, R, NIN, NON as the number of filters, input channels, input cols, intput rows, filter columns, filter rows, input neurons, output neurons.
We summarize the TC and DNN workloads that we use in the case studies in ~\autoref{table:tc_workloads} and ~\autoref{table:dnn_workloads}, respectively.

\subsection{Algorithm Exploration}

\begin{figure}[t]
\centerline{\includegraphics[scale=0.065]{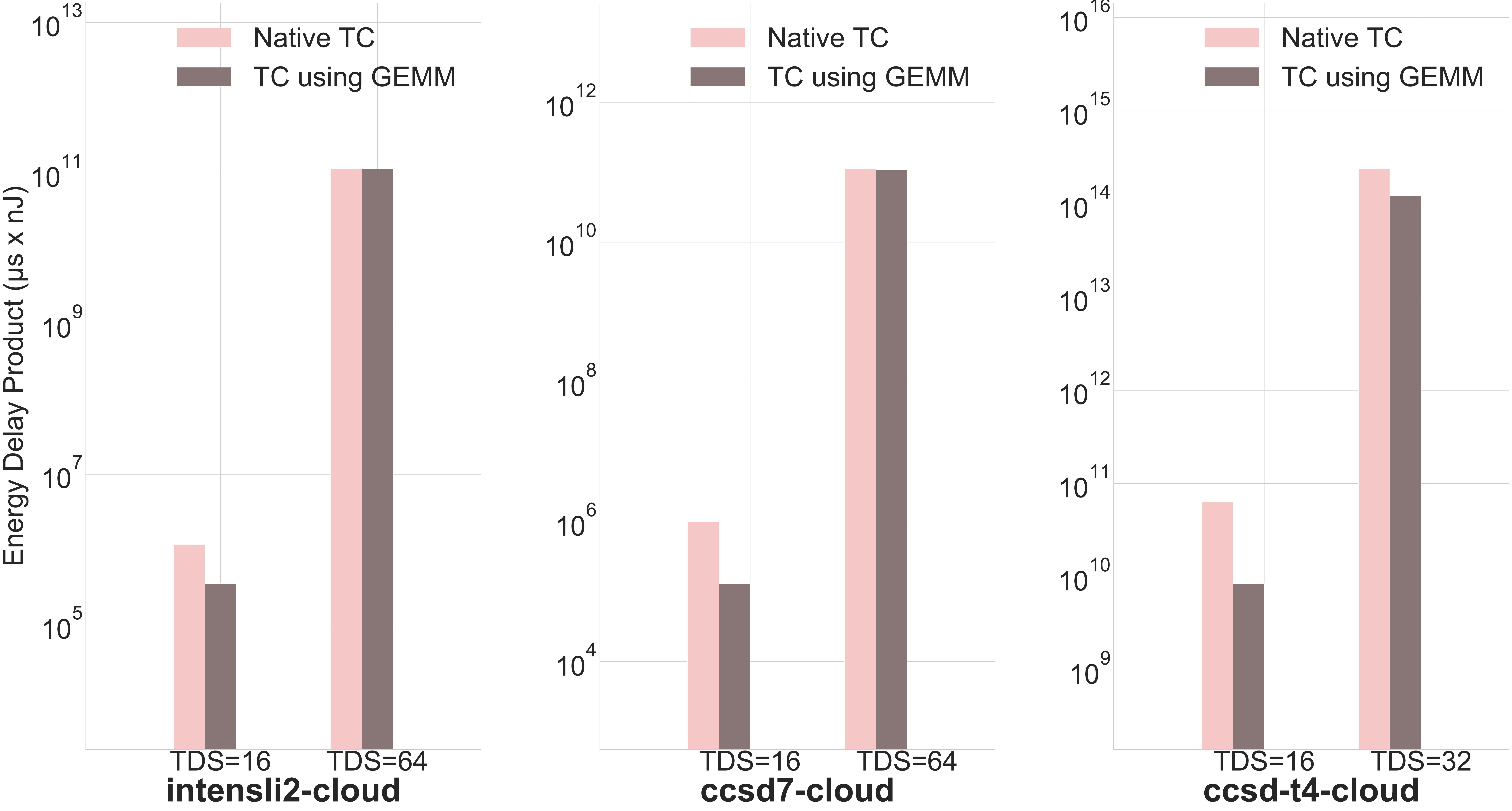}}
\caption{Three tensor contraction examples with different dimensions using different algorithms (native and TTGT) on a cloud accelerator. We explore Tensor Dimension Sizes (TDS) with 16 and 64 for intensli2 and ccsd7, and 16 and 32 for ccsd-t4.}
\label{fig:algorith-exploration}
\end{figure}

\begin{figure}[t]
\centerline{\includegraphics[scale=0.35]{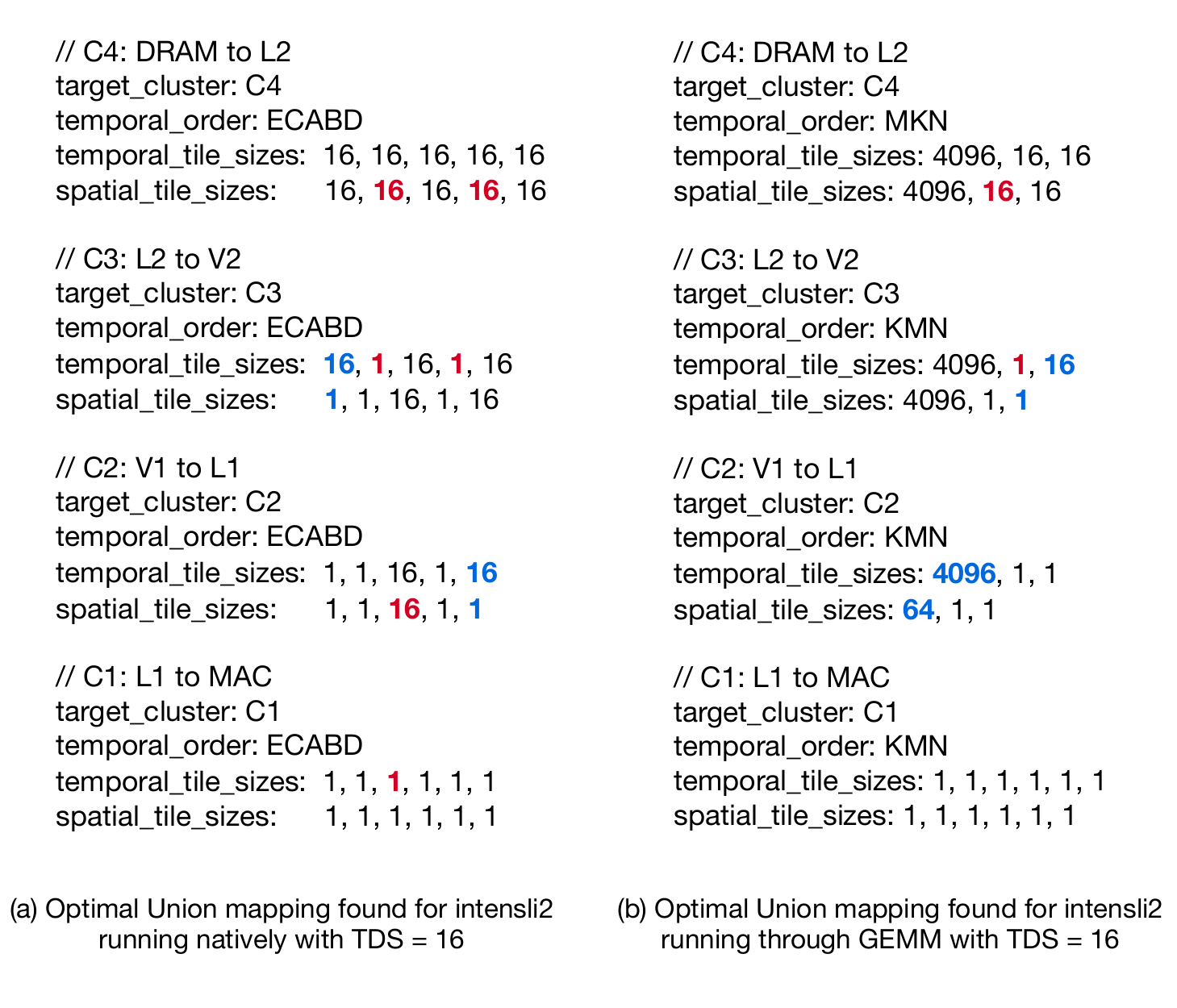}}
\caption{Generated mappings from \union for intensli2 using different algorithms with tensor dimension sizes as 16. The orders of dimensions in tile\_sizes are ABCDE and MNK for the mappings in (a) and (b) respectively. Blue tilesizes show the spatial distribution while red tilesizes show the temporal distribution for the dimension.}
\label{fig:intensli-mapping}
\end{figure}

A single tensor operation can be computed via several algorithms. The \union-frontend determines whether to run an operation
natively, or transform it 
to other operations, depending 
on which algorithm provides better performance on the 
accelerator.
We demonstrate 
this use case using 
tensor contraction running 
on a cloud type 2D spatial accelerator via  two algorithms: (1) running natively and (2) running through TTGT. 
We use the Timeloop cost model and a mapper based on both heuristic and random sampling.
We use the cloud configuration in ~\autoref{table:hardware_config} with 32$\times$64 as the aspect ratio of the accelerator to balance the parallelism across rows and columns.
%We explore how different mappings are need to achieve the best performance and energy efficiency.
%We ran both timeloop for both cases with 2D spatial architecture similar to ~\autoref{fig:union_interface}(c).
Note that for TTGT cost estimation, the cost model only estimates the cost of the GEMM operation assuming that the cost of transpose operations would not be significant.
%we only estimate the cost of GEMM operation and approximate the cost of operation as the GEMM operation.
Since TTGT does not incur duplicated elements of the original tensors, the memory footprint for both running TC natively and running TC with TTGT have the same memory footprint.
\autoref{fig:algorith-exploration} plots the Energy-Delay-Product (EDP) for three tensor contractions with tensor dimensions 16 and 64 on the cloud accelerator.
We observe that the lower EDP is achieved when running with TTGT for all cases with TDS=16.
This is because running natively will under-utilize the available compute units since the target accelerator has 32$\times$64 PEs while the each tensor dimension has size of 16. 
For example, \autoref{fig:intensli-mapping} shows the mappings generated from \union for Intensli2.
In ~\autoref{fig:intensli-mapping}(a) $C_3$ level, we observe that the optimal mapping found by \union distributes the problem dimension A across 16 $C_2$s and distributes the dimension E across 16 $C_1$s, resulting in utilizing 256 PEs with \textit{A\_E} partitioned mapping.
In ~\autoref{fig:intensli-mapping}(b), the optimal mapping with GEMM distributes K across 16 $C_2$s and distributes M across 64 $C_1$s, resulting in utilizing 1024 PEs with \textit{K\_M} partitioned mapping.

%Therefore, we observe that as long as the bandwidth is sufficient, temporal/spatial tiling is done reasonably, the performance should be similar as shown in the first row of ~\autoref{fig:algorith-exploration}.
%Since the number of rows and columns are small enough for the edge accelerators, all cases were able to fully exploit the available parallelism. 

\subsection{Mapping Exploration}

\begin{figure}[t]
\centerline{\includegraphics[scale=0.065]{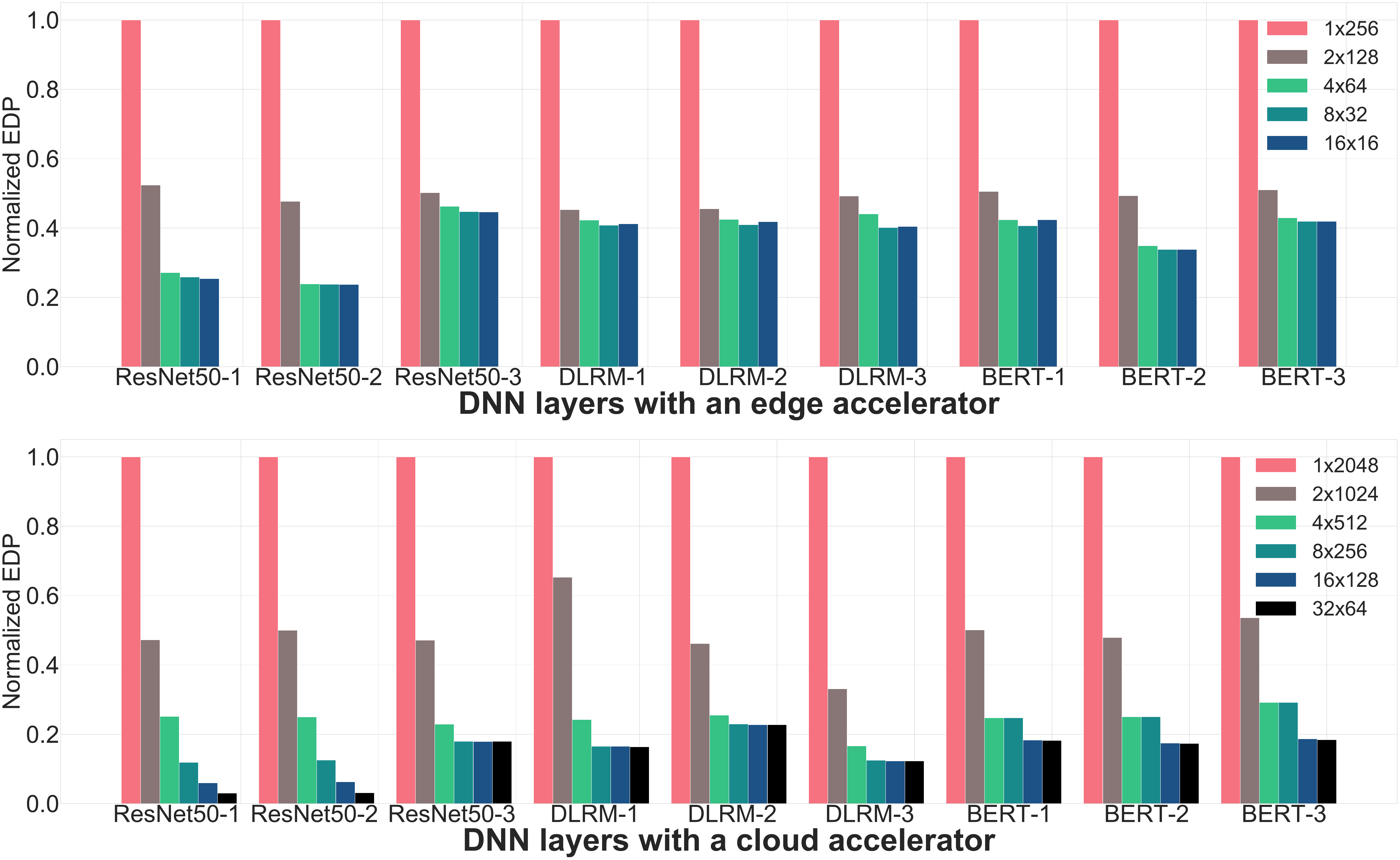}}
\caption{EDP comparison on DNN workloads using a flexible accelerator with different aspect ratio.}
\label{fig:aspect-ratio}
\end{figure}

Flexible accelerators like Eyeriss\_v2~\cite{eyeriss_v2} and MAERI~\cite{maeri_asplos18} can logically configure to different aspect ratios for the underlying PE array
via configurable NoCs.
This flexibility allows these accelerators to efficiently run layers with different shapes and sizes.
%by configuring Virtual Neurons with different sizes. 
We demonstrate the value of 
\union by exploring 
optimized array configurations for different DNN workloads
for such flexible substrates.
For this case study, we use the MAESTRO cost model as 
it has support to model such flexible accelerators.
The flexibility in aspect ratios gets captured 
by allowing cluster sizes to be variable.
In the \union constraint file, we specify different cluster sizes to explore different aspect ratio.
%
%To demonstrate the impact of aspect ratio for different worklaods,

We evaluate the DNN workloads shown in ~\autoref{table:dnn_workloads} using different aspect ratio for the edge (1$\times$256, 2$\times$128, 4$\times$64, 8$\times$32, and 16$\times$16) and cloud accelerators (1$\times$2048, 2$\times$1024, 4$\times$512, 8$\times$256, 16$\times$128, and 32$\times$64). Each aspect ratio corresponds to a configuration of the flexible accelerator.
%We use memory-centric approach which forces 1-to-1 mapping of problem dimension and spatial dimension (i.e. same problem dimension cannot be distributed into different spatial dimensions).
\autoref{fig:aspect-ratio} plots the EDP. 
We observe that the EDP gets saturated once it maximizes the PE utilization after the mapper finds the optimal tile sizes and loop orders to maximize the data reuse.
Even though the balanced aspect ratio showed the best performance for most of the cases that we evaluate, this can be sub-optimal if the workload is unbalanced. 
For example, GEMM with 4$\times$2048 or 2048$\times$2 or 4$\times$2 will be able to fully utilize an accelerator with 1$\times$2048 aspect ratio by parallelizing K dimension while 32$\times$64 accelerator will be underutilized.
%However, this can be overcome by using the 
This is where \union's
cluster-centric approach to describe mappings helps 
as it enables 
mapping the same workload dimensions to different spatial dimensions to fully exploit the available parallelism.
%There is no single aspect ratio which performs best for all different DNN workloads since the performance and energy consumption highly depend on various factors including problem types, shapes, aspect ratios and etc. 

% Performance goal is based on the number of PEs and the clock rate.
% We assume a 1 GHz clock for the accelerators.

%\tushar{a nice have would have been a follow-on case study here running two different mappers on same cost model. I wonder if we can pull some data out of Marvel paper / results for this case study just to show the impact of mappers in terms of their runtime and final found solution. That case study fits here in the mapping exploration for flexible accelerator subsection. @prasanth?}
%\prasanth{Sure, should we go for it? The catch is that the data we have is not for the layers in Table V.}

\subsection{Hardware Exploration}
\begin{figure}[t]
\centerline{\includegraphics[scale=0.068]{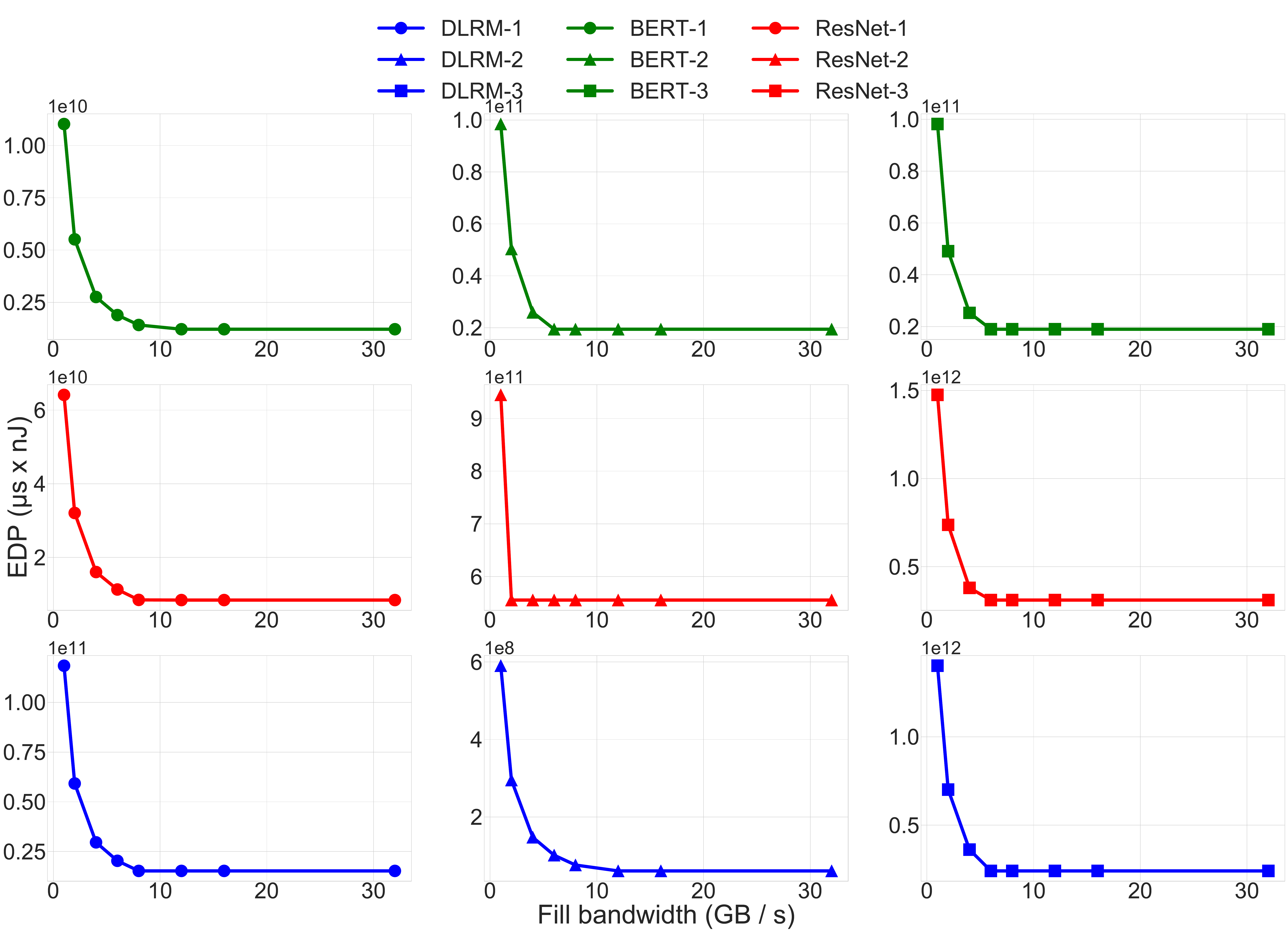}}
\caption{EDP comparison with different fill bandwidth on multi-chiplet based architecture using DNN workloads.}
\label{fig:sensitivity-fill-bw}
\end{figure}
In our last case study, 
we study the impact of chipletization on an accelerator's performance. 
%Given the recent trend towards chipletization to save chip manufacturing costs, we compare a monolithic single-chip accelerator versus a multi-chiplet one like Simba~\cite{simba_micro19}.
%Both accelerators have the same number of PEs.
Multi-chiplet based architectures are gaining popularity as they can reduce manufacturing cost and provide scalabililty. NVIDIA's Simba~\cite{simba_micro19} is a recent example.
%are used due to its modularity and scalabilty.
However, the inter-chiplet network is more expensive than on-chip network resulting in lower bandwidth and higher energy. 
For this case study, we use an accelerator which is composed of 16 chiplets, and each chiplet has the same configuration with the edge accelerator in ~\autoref{table:hardware_config}. 
The total number of PEs are equal to 4096.  
We study the impact of the interconnect bandwidth by varying the fill bandwidth of the global buffer in each chiplet, i.e. the bandwidth from DRAM to the global buffer in each chiplet. 
We use Timeloop for this case study as it can model hierarchical architectures like Simba and also comes with the Accelergy~\cite{accelergy_iccad19} energy model 
for accurately estimating on-chip versus on-package energy.

\autoref{fig:sensitivity-fill-bw} plots our results.
For all models, we observe that EDP drops rapidly with the increase in fill bandwidth when the fill bandwidth is low, and it gets saturated once the fill bandwidth is sufficient so that it is not bounded by the fill bandwidth.
According to the result, different layers get saturated in the different fill bandwidth depending on the available data reuse.
We also observe that ResNet-2 gets saturated when fill bandwidth is 2GB/s while others get saturated between 6~-12 GB/s.

\section{Conclusion}
In this work, we propose \union, a unified framework for evaluating tensor operations on spatial accelerators. Our MLIR based framework allows to map both HPC and ML tensor operations using multiple mappers to multiple cost models for spatial accelerators. The three case studies presented demonstrate the flexibility of the framework by evaluating very different operations, mappings, and hardware features with a single framework. While the number of operations, mappings and accelerators are currently limited to what we have demonstrated here, we plan to extend to other kernels such as tensor decomposition, other accelerators and mappers in the near future.
There are also advanced features that can be added to \union abstractions to support fused operations, sparsity-aware accelerator cost models, and unimodular/polyhedral mappings, but we leave those as a future work.
The modular framework allows us to add such changes without requiring a redesign of the whole software stack.

\section{Acknowledgment}
\label{sec:ack}
%\vspace{-1mm}
We thank Ruiqin Tian at PNNL for her help with the COMET compiler. We also thank Hyoukjun Kwon, Clayton Hughes, Mark Plagge, Juan Escobedo, Rizwan Ashraf for insightful comments and discussions on this work. 
We also thank the anonymous reviewers for their valuable feedback. 
Support for this work was provided through U.S. Department of Energy’s (DOE) Office of Advanced Scientific Computing Research as part of the Center for Artificial Intelligence-focused Architectures and Algorithms. PNNL is operated by Battelle for the DOE under Contract DE-AC05-76RL01830.
Sandia National Laboratories is a multimission laboratory managed and operated by National Technology and Engineering Solutions of Sandia, LLC., a  wholly owned subsidiary of Honeywell International, Inc., for the U.S. Department of Energy’s National Nuclear Security Administration under contract DE-NA-0003525.

\newpage

%Names should not be listed in columns nor group by affiliation. Please keep your affiliations as succinct as possible (for example, do not differentiate among departments of the same organization).

%Use the abbreviation ``Fig.~\ref{fig}'', even at the beginning of a sentence.

% \section*{Acknowledgment}
% The preferred spelling of the word ``acknowledgment'' in America is without 
% an ``e'' after the ``g''. Avoid the stilted expression ``one of us (R. B. 
% G.) thanks $\ldots$''. Instead, try ``R. B. G. thanks$\ldots$''. Put sponsor 
% acknowledgments in the unnumbered footnote on the first page.

\bibliographystyle{IEEEtran}
\bibliography{references}

\begin{thebibliography}{10}
\providecommand{\url}[1]{#1}
\csname url@samestyle\endcsname
\providecommand{\newblock}{\relax}
\providecommand{\bibinfo}[2]{#2}
\providecommand{\BIBentrySTDinterwordspacing}{\spaceskip=0pt\relax}
\providecommand{\BIBentryALTinterwordstretchfactor}{4}
\providecommand{\BIBentryALTinterwordspacing}{\spaceskip=\fontdimen2\font plus
\BIBentryALTinterwordstretchfactor\fontdimen3\font minus
  \fontdimen4\font\relax}
\providecommand{\BIBforeignlanguage}[2]{{%
\expandafter\ifx\csname l@#1\endcsname\relax
\typeout{** WARNING: IEEEtran.bst: No hyphenation pattern has been}%
\typeout{** loaded for the language `#1'. Using the pattern for}%
\typeout{** the default language instead.}%
\else
\language=\csname l@#1\endcsname
\fi
#2}}
\providecommand{\BIBdecl}{\relax}
\BIBdecl

\bibitem{tpu_isca17}
N.~P. Jouppi, C.~Young, N.~Patil, D.~Patterson, G.~Agrawal, R.~Bajwa, S.~Bates,
  S.~Bhatia, N.~Boden, A.~Borchers, R.~Boyle, P.-l. Cantin, C.~Chao, C.~Clark,
  J.~Coriell, M.~Daley, M.~Dau, J.~Dean, B.~Gelb, T.~V. Ghaemmaghami,
  R.~Gottipati, W.~Gulland, R.~Hagmann, C.~R. Ho, D.~Hogberg, J.~Hu, R.~Hundt,
  D.~Hurt, J.~Ibarz, A.~Jaffey, A.~Jaworski, A.~Kaplan, H.~Khaitan,
  D.~Killebrew, A.~Koch, N.~Kumar, S.~Lacy, J.~Laudon, J.~Law, D.~Le, C.~Leary,
  Z.~Liu, K.~Lucke, A.~Lundin, G.~MacKean, A.~Maggiore, M.~Mahony, K.~Miller,
  R.~Nagarajan, R.~Narayanaswami, R.~Ni, K.~Nix, T.~Norrie, M.~Omernick,
  N.~Penukonda, A.~Phelps, J.~Ross, M.~Ross, A.~Salek, E.~Samadiani, C.~Severn,
  G.~Sizikov, M.~Snelham, J.~Souter, D.~Steinberg, A.~Swing, M.~Tan,
  G.~Thorson, B.~Tian, H.~Toma, E.~Tuttle, V.~Vasudevan, R.~Walter, W.~Wang,
  E.~Wilcox, and D.~H. Yoon, ``In-datacenter performance analysis of a tensor
  processing unit,'' in \emph{Proceedings of the 44th Annual International
  Symposium on Computer Architecture}.\hskip 1em plus 0.5em minus 0.4em\relax
  New York, NY, USA: Association for Computing Machinery, 2017, p. 1–12.

\bibitem{xDNN-web}
``Accelerating dnns with xilinx alveo accelerator cards,''
  https://www.xilinx.com/support/documentation/white\_papers/wp504-accel-dnns.pdf.

\bibitem{DBLP:conf/vlsic/FleischerSZSOSC18}
B.~M. Fleischer \emph{et~al.}, ``{A Scalable Multi- TeraOPS Deep Learning
  Processor Core for {AI} Trainina and Inference},'' in \emph{2018 {IEEE}
  Symposium on {VLSI} Circuits}, 2018, pp. 35--36.

\bibitem{nvdla}
NVIDIA, ``The nvidia deep learning accelerator (nvdla),''
  \url{http://nvdla.org/hw/v1/ias/programming_guide.html}.

\bibitem{eyeriss_isca16}
Y.-H. Chen, J.~Emer, and V.~Sze, ``Eyeriss: A spatial architecture for
  energy-efficient dataflow for convolutional neural networks,'' in
  \emph{Proceedings of the 43rd International Symposium on Computer
  Architecture}, 2016, p. 367–379.

\bibitem{shidiannao_isca15}
Z.~{Du}, R.~{Fasthuber}, T.~{Chen}, P.~{Ienne}, L.~{Li}, T.~{Luo}, X.~{Feng},
  Y.~{Chen}, and O.~{Temam}, ``Shidiannao: Shifting vision processing closer to
  the sensor,'' in \emph{2015 ACM/IEEE 42nd Annual International Symposium on
  Computer Architecture (ISCA)}, 2015, pp. 92--104.

\bibitem{maeri_asplos18}
H.~Kwon, A.~Samajdar, and T.~Krishna, ``Maeri: Enabling flexible dataflow
  mapping over dnn accelerators via reconfigurable interconnects,'' in
  \emph{Proceedings of the Twenty-Third International Conference on
  Architectural Support for Programming Languages and Operating Systems
  (ASPLOS)}.\hskip 1em plus 0.5em minus 0.4em\relax New York, NY, USA:
  Association for Computing Machinery, 2018.

\bibitem{scalesim_ispass20}
A.~{Samajdar}, J.~M. {Joseph}, Y.~{Zhu}, P.~{Whatmough}, M.~{Mattina}, and
  T.~{Krishna}, ``A systematic methodology for characterizing scalability of
  dnn accelerators using scale-sim,'' in \emph{2020 IEEE International
  Symposium on Performance Analysis of Systems and Software (ISPASS)}, 2020,
  pp. 58--68.

\bibitem{stonne_20}
F.~{Mu{\~n}oz-Mart{\'\i}nez}, J.~L. {Abell{\'a}n}, M.~E. {Acacio}, and
  T.~{Krishna}, ``{STONNE: A Detailed Architectural Simulator for Flexible
  Neural Network Accelerators},'' \emph{arXiv e-prints}, p. arXiv:2006.07137,
  Jun. 2020.

\bibitem{maestro_micro19}
H.~Kwon, P.~Chatarasi, M.~Pellauer, A.~Parashar, V.~Sarkar, and T.~Krishna,
  ``Understanding reuse, performance, and hardware cost of dnn dataflow: A
  data-centric approach,'' in \emph{MICRO}, 2019.

\bibitem{timeloop_ispass19}
A.~{Parashar}, P.~{Raina}, Y.~S. {Shao}, Y.~{Chen}, V.~A. {Ying}, A.~{Mukkara},
  R.~{Venkatesan}, B.~{Khailany}, S.~W. {Keckler}, and J.~{Emer}, ``Timeloop: A
  systematic approach to dnn accelerator evaluation,'' in \emph{2019 IEEE
  International Symposium on Performance Analysis of Systems and Software
  (ISPASS)}, 2019, pp. 304--315.

\bibitem{interstellar_asplos20}
X.~Yang, M.~Gao, Q.~Liu, J.~Setter, J.~Pu, A.~Nayak, S.~Bell, K.~Cao, H.~Ha,
  P.~Raina, C.~Kozyrakis, and M.~Horowitz, ``Interstellar: Using halide's
  scheduling language to analyze dnn accelerators,'' in \emph{Proceedings of
  the Twenty-Fifth International Conference on Architectural Support for
  Programming Languages and Operating Systems (ASPLOS)}, 2020.

\bibitem{marvel}
P.~Chatarasi, H.~Kwon, N.~Raina, S.~Malik, V.~Haridas, A.~Parashar,
  M.~Pellauer, T.~Krishna, and V.~Sarkar, ``Marvel: A data-centric compiler for
  dnn operators on spatial accelerators,'' \emph{arXiv preprint
  arXiv:2002.07752}, 2020.

\bibitem{dmazerunner}
S.~Dave, Y.~Kim, S.~Avancha, K.~Lee, and A.~Shrivastava, ``Dmazerunner:
  Executing perfectly nested loops on dataflow accelerators,'' \emph{ACM Trans.
  Embed. Comput. Syst.}, Oct. 2019.

\bibitem{gamma_iccad20}
S.-C. Kao and T.~Krishna, ``Gamma: Automating the hw mapping of dnn models on
  accelerators via genetic algorithm,'' in \emph{Proceedings of the 39th
  International Conference on Computer-Aided Design (ICCAD)}, 2020.

\bibitem{Flash2021}
G.~E. Moon, H.~Kwon, G.~Jeong, P.~Chatarasi, S.~Rajamanickam, and T.~Krishna,
  ``Evaluating spatial accelerator architectures with tiled matrix-matrix
  multiplication,'' \emph{IEEE Transactions on Parallel and Distributed Systems
  (TPDS)}, 2021.

\bibitem{loma2021}
A.~Symons, L.~Mei, and M.~Verhelst, ``Loma: Fast auto-scheduling on dnn
  accelerators through loop-order-based memory allocation,'' in \emph{2021 IEEE
  3rd International Conference on Artificial Intelligence Circuits and Systems
  (AICAS)}, 2021, pp. 1--4.

\bibitem{simba_micro19}
Y.~S. Shao, J.~Clemons, R.~Venkatesan, B.~Zimmer, M.~Fojtik, N.~Jiang,
  B.~Keller, A.~Klinefelter, N.~Pinckney, P.~Raina \emph{et~al.}, ``Simba:
  Scaling deep-learning inference with multi-chip-module-based architecture,''
  in \emph{Proceedings of the 52nd Annual IEEE/ACM International Symposium on
  Microarchitecture}, 2019, pp. 14--27.

\bibitem{eyeriss_v2}
Y.-H. Chen, T.-J. Yang, J.~Emer, and V.~Sze, ``Eyeriss v2: A flexible
  accelerator for emerging deep neural networks on mobile devices,'' \emph{IEEE
  Journal on Emerging and Selected Topics in Circuits and Systems}, vol.~9,
  no.~2, pp. 292--308, 2019.

\bibitem{buffets_asplos19}
M.~Pellauer, Y.~S. Shao, J.~Clemons, N.~Crago, K.~Hegde, R.~Venkatesan, S.~W.
  Keckler, C.~W. Fletcher, and J.~Emer, ``Buffets: An efficient and composable
  storage idiom for explicit decoupled data orchestration,'' in
  \emph{Proceedings of the Twenty-Fourth International Conference on
  Architectural Support for Programming Languages and Operating Systems}, 2019,
  pp. 137--151.

\bibitem{mlir_cgo21}
C.~Lattner, M.~Amini, U.~Bondhugula, A.~Cohen, A.~Davis, J.~Pienaar, R.~Riddle,
  T.~Shpeisman, N.~Vasilache, and O.~Zinenko, ``Mlir: Scaling compiler
  infrastructure for domain specific computation,'' in \emph{CGO}, 2021.

\bibitem{mutlu2020comet}
E.~Mutlu, R.~Tian, B.~Ren, S.~Krishnamoorthy, R.~Gioiosa, J.~Pienaar, and
  G.~Kestor, ``Comet: A domain-specific compilation of high-performance
  computational chemistry,'' in \emph{Workshop on Languages and Compilers for
  Parallel Computing (LCPC'20)}.\hskip 1em plus 0.5em minus 0.4em\relax
  Springer.

\bibitem{autosa_fpga21}
J.~Wang, L.~Guo, and J.~Cong, ``Autosa: A polyhedral compiler for
  high-performance systolic arrays on fpga,'' in \emph{The 2021 ACM/SIGDA
  International Symposium on Field-Programmable Gate Arrays}, ser. FPGA '21,
  2021, p. 93–104.

\bibitem{zigzag2021}
L.~Mei, P.~Houshmand, V.~Jain, S.~Giraldo, and M.~Verhelst, ``Zigzag: Enlarging
  joint architecture-mapping design space exploration for dnn accelerators,''
  \emph{IEEE Transactions on Computers}, vol.~70, no.~8, pp. 1160--1174, 2021.

\bibitem{tvm_osdi18}
T.~Chen, T.~Moreau, Z.~Jiang, L.~Zheng, E.~Yan, M.~Cowan, H.~Shen, L.~Wang,
  Y.~Hu, L.~Ceze, C.~Guestrin, and A.~Krishnamurthy, ``Tvm: An automated
  end-to-end optimizing compiler for deep learning,'' in \emph{Proceedings of
  the 13th USENIX Conference on Operating Systems Design and Implementation},
  ser. OSDI'18, 2018, p. 579–594.

\bibitem{cudnn}
S.~Chetlur, C.~Woolley, P.~Vandermersch, J.~Cohen, J.~Tran, B.~Catanzaro, and
  E.~Shelhamer, ``cudnn: Efficient primitives for deep learning,'' \emph{arXiv
  preprint arXiv:1410.0759}, 2014.

\bibitem{ccsd_t}
K.~Raghavachari, G.~Trucks, J.~A.~Pople, and M.~Head-Gordon, ``A fifth-order
  perturbation comparison of electron correlation theories,'' \emph{Chemical
  Physics Letters}, vol. 157, pp. 479--483, 05 1989.

\bibitem{nwchem}
E.~Apr\`a, E.~J. Bylaska, W.~A. de~Jong, N.~Govind, K.~Kowalski, T.~P.
  Straatsma, M.~Valiev, H.~J.~J. van Dam, Y.~Alexeev, J.~Anchell, V.~Anisimov,
  F.~W. Aquino, R.~Atta-Fynn, J.~Autschbach, N.~P. Bauman, J.~C. Becca, D.~E.
  Bernholdt, K.~Bhaskaran-Nair, S.~Bogatko, P.~Borowski, J.~Boschen, J.~Brabec,
  A.~Bruner, E.~Cau\"et, Y.~Chen, G.~N. Chuev, C.~J. Cramer, J.~Daily, M.~J.~O.
  Deegan, T.~H.~D. Jr., M.~Dupuis, K.~G. Dyall, G.~I. Fann, S.~A. Fischer,
  A.~Fonari, H.~Fr\"uchtl, L.~Gagliardi, J.~Garza, N.~Gawande, S.~Ghosh,
  K.~Glaesemann, A.~W. G\"{o}tz, J.~Hammond, V.~Helms, E.~D. Hermes, K.~Hirao,
  S.~Hirata, M.~Jacquelin, L.~Jensen, B.~G. Johnson, H.~J{\'o}nsson, R.~A.
  Kendall, M.~Klemm, R.~Kobayashi, V.~Konkov, S.~Krishnamoorthy, M.~Krishnan,
  Z.~Lin, R.~D. Lins, R.~J. Littlefield, A.~J. Logsdail, K.~Lopata, W.~Ma,
  A.~V. Marenich, J.~M. del Campo, D.~Mejia-Rodriguez, J.~E. Moore, J.~M.
  Mullin, T.~Nakajima, D.~R. Nascimento, J.~A. Nichols, P.~J. Nichols,
  J.~Nieplocha, A.~O. de~la Roza, B.~Palmer, A.~Panyala, T.~Pirojsirikul,
  B.~Peng, R.~Peverati, J.~Pittner, L.~Pollack, R.~M. Richard, P.~Sadayappan,
  G.~C. Schatz, W.~A. Shelton, D.~W. Silverstein, D.~M.~A. Smith, T.~A. Soares,
  D.~Song, M.~Swart, H.~L. Taylor, G.~S. Thomas, V.~Tipparaju, D.~G. Truhlar,
  K.~Tsemekhman, T.~V. Voorhis, A.~V\'azquez-Mayagoitia, P.~Verma, O.~Villa,
  A.~Vishnu, K.~D. Vogiatzis, D.~Wang, J.~H. Weare, M.~J. Williamson, T.~L.
  Windus, K.~Woli\'{n}ski, A.~T. Wong, Q.~Wu, C.~Yang, Q.~Yu, M.~Zacharias,
  Z.~Zhang, Y.~Zhao, and R.~J. Harrison, ``{NWChem: Past, Present, and
  Future},'' \emph{Journal of Chemical Physics}, vol. 152, no.~17, p. 184102,
  May 2020.

\bibitem{nwchemex_chemrev2021}
K.~Kowalski, R.~Bair, N.~P. Bauman, J.~S. Boschen, E.~J. Bylaska, J.~Daily,
  W.~A. de~Jong, D.~T.~H. Jr., N.~Govind, R.~J. Harrison, M.~Ke\c{c}eli,
  K.~Keipert, S.~Krishnamoorthy, S.~Kumar, E.~Mutlu, B.~Palmer, A.~Panyala,
  B.~Peng, R.~M. Richard, T.~P. Straatsma, P.~Sushko, E.~F. Valeev, M.~Valiev,
  H.~J.~J. van Dam, J.~M. Waldrop, D.~B. Williams-Young, C.~Yang, M.~Zalewski,
  and T.~L. Windus, ``{From NWChem to NWChemEx: Evolving with the computational
  chemistry landscape},'' \emph{Chemical Reviews}, accepted for publication.

\bibitem{tensorcore}
``Nvidia tesla v100 gpu architecture.'' 2017,
  \url{https://images.nvidia.com/content/volta-architecture/pdf/volta-architecture-whitepaper.pdf}.

\bibitem{sigma_hpca20}
E.~{Qin}, A.~{Samajdar}, H.~{Kwon}, V.~{Nadella}, S.~{Srinivasan}, D.~{Das},
  B.~{Kaul}, and T.~{Krishna}, ``Sigma: A sparse and irregular gemm accelerator
  with flexible interconnects for dnn training,'' in \emph{2020 IEEE
  International Symposium on High Performance Computer Architecture (HPCA)},
  2020, pp. 58--70.

\bibitem{tetris_asplos17}
M.~Gao, J.~Pu, X.~Yang, M.~Horowitz, and C.~Kozyrakis, ``Tetris: Scalable and
  efficient neural network acceleration with 3d memory,'' in \emph{Proceedings
  of the Twenty-Second International Conference on Architectural Support for
  Programming Languages and Operating Systems}, 2017, pp. 751--764.

\bibitem{mindmapping_asplos21}
K.~Hegde, P.-A. Tsai, S.~Huang, V.~Chandra, A.~Parashar, and C.~W. Fletcher,
  ``Mind mappings: Enabling efficient algorithm-accelerator mapping space
  search,'' in \emph{Proceedings of the Twenty-Fifth International Conference
  on Architectural Support for Programming Languages and Operating Systems
  (ASPLOS)}, 2021.

\bibitem{sze2020evaluate}
V.~Sze, Y.-H. Chen, T.-J. Yang, and J.~S. Emer, ``How to evaluate deep neural
  network processors: Tops/w (alone) considered harmful,'' \emph{IEEE
  Solid-State Circuits Magazine}, vol.~12, no.~3, pp. 28--41, 2020.

\bibitem{mlir-iree}
``{IREE}: Intermediate representation execution environment,'' 2021,
  \url{https://github.com/google/iree}.

\bibitem{mlir-npcomp}
``{NPComp} - mlir based compiler toolkit for numerical python programs,'' 2021,
  \url{https://github.com/llvm/mlir-npcomp}.

\bibitem{mlir-tosa}
``Tensor operator set architecture (tosa) dialect,'' 2021,
  \url{https://mlir.llvm.org/docs/Dialects/TOSA/}.

\bibitem{tian2021sparseComet}
R.~Tian, L.~Guo, J.~Li, B.~Ren, and G.~Kestor, ``A high-performance sparse
  tensor algebra compiler in {Multi-Level IR},'' \emph{arXiv preprint
  arXiv:2102.05187}, 2021.

\bibitem{dlrm}
\BIBentryALTinterwordspacing
M.~Naumov, D.~Mudigere, H.~M. Shi, J.~Huang, N.~Sundaraman, J.~Park, X.~Wang,
  U.~Gupta, C.~Wu, A.~G. Azzolini, D.~Dzhulgakov, A.~Mallevich,
  I.~Cherniavskii, Y.~Lu, R.~Krishnamoorthi, A.~Yu, V.~Kondratenko, S.~Pereira,
  X.~Chen, W.~Chen, V.~Rao, B.~Jia, L.~Xiong, and M.~Smelyanskiy, ``Deep
  learning recommendation model for personalization and recommendation
  systems,'' \emph{CoRR}, vol. abs/1906.00091, 2019. [Online]. Available:
  \url{http://arxiv.org/abs/1906.00091}
\BIBentrySTDinterwordspacing

\bibitem{SpringerTCCG2018}
P.~Springer and P.~Bientinesi, ``Design of a high-performance gemm-like
  tensor–tensor multiplication,'' \emph{ACM Trans. Math. Softw.}, vol.~44,
  no.~3, 2018.

\bibitem{accelergy_iccad19}
Y.~N. Wu, J.~S. Emer, and V.~Sze, ``Accelergy: An architecture-level energy
  estimation methodology for accelerator designs,'' in \emph{2019 IEEE/ACM
  International Conference on Computer-Aided Design (ICCAD)}.\hskip 1em plus
  0.5em minus 0.4em\relax IEEE, 2019, pp. 1--8.

\end{thebibliography}

%\newpage
%\clearpage
%\newpage
%\input{pact20_latex_template/appendix}
\end{document}